\documentclass[%
 reprint,
superscriptaddress,
 amsmath,amssymb,
 aps,
]{revtex4-2}

\usepackage{graphicx}
\usepackage{dcolumn}
\usepackage{bm}
\usepackage{hyperref}

\usepackage{todonotes,booktabs}

\begin{document}

\title{Spectroscopy on the eEDM-sensitive states of ThF$^+$}

\author{Kia Boon Ng}
    \affiliation{JILA, NIST and University of Colorado, and Department of Physics, University of Colorado, Boulder CO 80309, USA.}
    \email{kia.ng@colorado.edu}

\author{Yan Zhou}
    \affiliation{JILA, NIST and University of Colorado, and Department of Physics, University of Colorado, Boulder CO 80309, USA.}
    \affiliation{Department of Physics and Astronomy, University of Nevada, Las Vegas, Las Vegas, NV 89154, USA.}

\author{Lan Cheng}
    \affiliation{Department of Chemistry, Johns Hopkins University, Baltimore, MD 21218, USA.}

\author{Noah Schlossberger}
    \affiliation{JILA, NIST and University of Colorado, and Department of Physics, University of Colorado, Boulder CO 80309, USA.}

\author{Sun Yool Park}
    \affiliation{JILA, NIST and University of Colorado, and Department of Physics, University of Colorado, Boulder CO 80309, USA.}
    
    

\author{Tanya S. Roussy}
    \affiliation{JILA, NIST and University of Colorado, and Department of Physics, University of Colorado, Boulder CO 80309, USA.}
    
    
\author{Luke Caldwell}
    \affiliation{JILA, NIST and University of Colorado, and Department of Physics, University of Colorado, Boulder CO 80309, USA.}
    
\author{Yuval Shagam}
    \affiliation{JILA, NIST and University of Colorado, and Department of Physics, University of Colorado, Boulder CO 80309, USA.}
    \affiliation{Schulich Faculty of Chemistry and Solid State Institute, Technion - Israel Institute of Technology, Haifa 3200003, Israel}

\author{Antonio J. Vigil}
    \affiliation{JILA, NIST and University of Colorado, and Department of Physics, University of Colorado, Boulder CO 80309, USA.}
    

\author{Eric A. Cornell}
    \affiliation{JILA, NIST and University of Colorado, and Department of Physics, University of Colorado, Boulder CO 80309, USA.}

\author{Jun Ye}
    \affiliation{JILA, NIST and University of Colorado, and Department of Physics, University of Colorado, Boulder CO 80309, USA.}

\date{\today}

\begin{abstract}
An excellent candidate molecule for the measurement of the electron's electric dipole moment (eEDM) is thorium monofluoride (ThF$^+$) because the eEDM-sensitive state, $^3\Delta_1$, is the electronic ground state, and thus is immune to decoherence from spontaneous decay. We perform spectroscopy on $X\,^3\Delta_1$ to extract three spectroscopic constants crucial to the eEDM experiment: the hyperfine coupling constant, the molecular frame electric dipole moment, and the magnetic $g$-factor. To understand the impact of thermal blackbody radiation on the vibrational ground state, we study the lifetime of the first excited vibrational manifold of $X\,^3\Delta_1$. We perform \textit{ab initio} calculations, compare them to our results, and discuss prospects for using ThF$^+$ in a new eEDM experiment at JILA.
\end{abstract}

\maketitle

\section{Introduction}
    The electron's electric dipole moment (eEDM) is strongly linked to our understanding of the universe \cite{hinds1997testing,khriplovich2012cp,Chupp2015,Stadnik2018,cesarotti2019interpreting}. One of the most successful models we have to describe the universe is the Standard Model of particle physics, and yet it is known to be incomplete. There has been a substantial effort on the theoretical front to introduce new physics through extensions of Standard Model \cite{sarkar1996big,ellis2007beyond}. These new physics models make varying predictions for the values of the eEDM \cite{pospelov1991electric,barr1990electric,bernreuther1991electric,barr1993review,pospelov2005electric,commins1999electric}. A measurement of (or an improved limit on) the eEDM would place constraints on these new theories.
    
    Sensitivity to the eEDM depends on three main factors in the experiment: (i) effective electric field strength that couples to the eEDM, (ii) coherent interrogation time of the eEDM-sensitive state, and (iii) total number of counts in the experiment for statistics. Groups with the world's best limits on the eEDM \cite{cairncross2017precision,acme2018improved,hudson2011improved} take advantage of the large effective electric field \cite{meyer2008prospects,Denis2015,skripnikov2015theoretical,petrov2007theoretical,leanhardt2011high,skripnikov2013communication,skripnikov2015theoreticalThO} in molecules to enhance their eEDM sensitivities. One of the ingredients for success in ACME \cite{acme2018improved,baron2014order} and Imperial College \cite{hudson2011improved} experiments is the large number of molecules that they probe in their neutral molecular beam experiments. On the other hand, our recent eEDM measurement at JILA \cite{cairncross2017precision} takes advantage of the long ion trapping times to tap the long coherence times of the eEDM-sensitive state. At present, we are enhancing our sensitivity through improvements in the trap design to accommodate more ions in addition to innovations for common-mode noise rejection \cite{zhou2020second}. New eEDM measurements with the improved setup are in progress, with the results due soon.
    
    Looking beyond our in-progress measurement, we plan to replace the molecule of choice, HfF$^+$, with $^{232}$ThF$^+$. The latter keeps all the benefits of using molecular ions in an ion trap, and it also boasts a larger effective electric field and longer coherence times than the former \cite{gresh2016broadband,Denis2015,skripnikov2015theoretical,petrov2007theoretical,leanhardt2011high}, both of which promise a direct increase in the eEDM sensitivity. Previous spectroscopic work \cite{zhou2019visible,heaven2014spectroscopy,barker2012spectroscopic,gresh2016broadband} shows that we can use similar experimental techniques across both molecular species, including multi-state detection \cite{zhou2020second,shagam2020continuous}. Hence, the molecule switch presents no new immediate experimental complexity, and promises higher eEDM sensitivity.
    
    Borrowing wisdom and experimental techniques from similar spectroscopic work performed on HfF$^+$ \cite{loh2011laser,loh2012rempi,ni2014state,loh2013precision}, ThF \cite{zhou2019visible}, and ThF$^+$ \cite{gresh2016broadband}, we (i) perform spectroscopy on the eEDM-sensitive state in ThF$^+$, $X\,^3\Delta_1$, (Section \ref{sec:spectroscopy}) to extract spectroscopic constants of concern, and discuss theoretical calculations of aforementioned spectroscopic constants; and (ii) study the lifetime of the first vibrational excited state in ThF$^+$ (Section \ref{sec:BBR}) and its implications on the expected coherence time of $X\,^3\Delta_1$.
    
\section[eEDM-sensitive state Spectroscopy]{\texorpdfstring{\MakeLowercase{e}}{e}EDM-sensitive state Spectroscopy}\label{sec:spectroscopy}
    Using ThF$^+$ in our eEDM experiment requires knowledge of the details of $X\,^3\Delta_1$, namely its responses to external electric and magnetic fields, and the frequencies of lasers required for state preparation and detection. We thus need to determine certain spectroscopic constants in our effective Hamiltonian governing our system. The effective Hamiltonian is very involved, and the interested reader is encouraged to consult our previous publication (Supplementary Material of Ref.\ \onlinecite{cairncross2017precision}) for more details. We shall introduce the relevant constants in context, below.
    
    The following sections detail the spectroscopy process to extract the spectroscopic constants crucial to the eEDM experiment. We begin with an overview of the state preparation process in Section \ref{sec:setup}, then we touch on the details of the measurement of the hyperfine coupling constant (Section \ref{sec:microwave_spectroscopy}), molecular frame electric dipole moment (Section \ref{sec:microwave_spectroscopy}), and the magnetic $g$-factor (Section \ref{sec:Ramsey}). We compare our experimental values with theoretical calculations in Section \ref{sec:theoreticalCalc}. Finally we present a summary of the results in Section \ref{sec:summary_spectroscopy}.
    
    \subsection{State preparation and readout for eEDM-sensitive state spectroscopy}\label{sec:setup}
\begin{figure}
            \centering
            \includegraphics[width=\columnwidth]{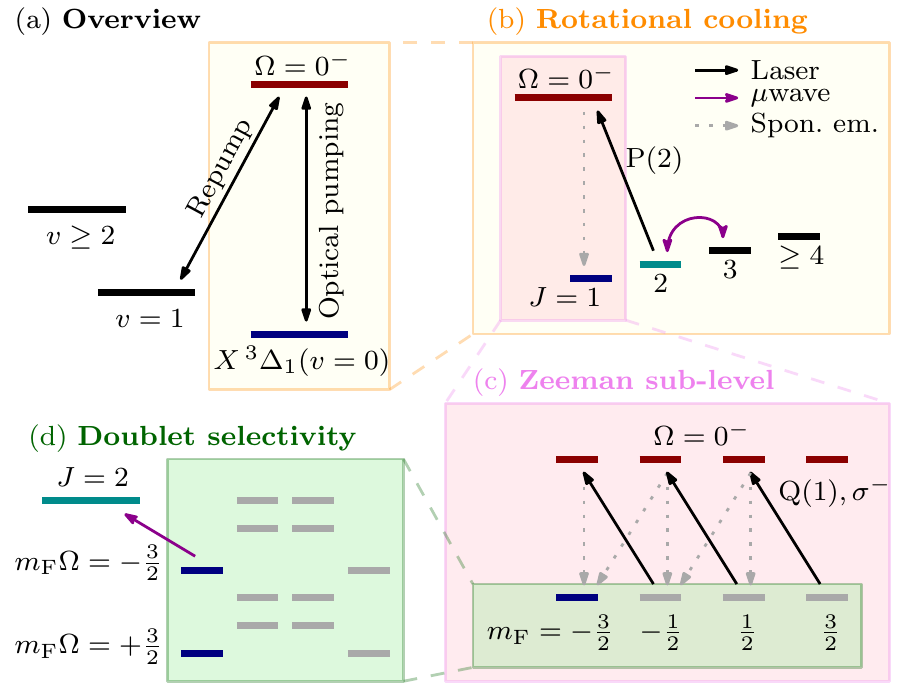}
            \caption{\textbf{State preparation sequence (not to scale).} (a) All our optical pumping lasers connect from the $X\,^3\Delta_1(v=0)$ vibronic manifold to the $\Omega=0^-(v=0)$ vibrational manifold \cite{gresh2016broadband}. We have a vibrational repump laser from the $X\,^3\Delta_1(v=1)$ manifold. (b) Within the $X\,^3\Delta_1 (v=0)$ vibronic manifold, we perform rotational cooling by optically pumping the $J=2$ state through the $\Omega=0^-$ state. We use microwaves to couple the $J=2$ and $J=3$ states together to transfer $J=3$ population eventually into $J=1$. (c) Within the $J=1$ manifold, we pump all the ions into a single Zeeman $m_F$ state using circularly polarized light on the Q(1) line. We also introduce a magnetic field to prevent the Zeeman levels from mixing through rotation coupling (refer to Section \ref{sec:Ramsey}). (d) $\Omega$-doubling gives rise to two closely spaced states with the same $m_F$ number. We can deplete one of these states by coupling it with microwaves to $J=2$. The state preparation process is shown as a sequence of steps for clarity, but all the steps involved occur at the same time in our experiment.}
            \label{fig:StatePreparation}
        \end{figure}

        The energy level of a diatomic molecule like ThF$^+$ has nested progressively finer structure. The eEDM-sensitive states are labeled by the quantum numbers $X\,^3\Delta_1 (v=0, J=1, F=3/2, m_F = \pm 3/2)$, which correspond to the electronic, vibrational, rotational, hyperfine, and Zeeman manifolds, respectively, with increasing fineness in their structures. We first prepare ThF$^+$ in the $X\,^3\Delta_1 (v=0)$ vibrational manifold through resonance-enhanced--multi-photon ionization of neutral ThF \cite{zhou2019visible}, and usher them into the finer energy structures with optical pumping via an excited electronic state $\Omega=0^-$ (Figure \ref{fig:StatePreparation}). The $\Omega=0^-$ electronic state lies approximately 14600~cm$^{-1}$ above $X\,^3\Delta_1$ \cite{gresh2016broadband}. The full process to prepare our ions in the $X\,^3\Delta_1 (v=0, J=1, F=3/2, m_F = -3/2, \Omega=1 )$ is illustrated in Figure \ref{fig:StatePreparation}, where $\Omega$ is the quantum number for the $\Omega$-doublets.
        
        State preparation involves two pulsed lasers at 304~nm and 532~nm for resonance-enhanced--multi-photon ionization, multiple cw lasers at 685~nm for optical pumping [Figures \ref{fig:StatePreparation}(a-c)], a cw repump laser at 717~nm [Figure \ref{fig:StatePreparation}(a)], and microwave channels at 29~GHz and 43~GHz [Figure \ref{fig:StatePreparation}(b,d)].
        
        We perform our state readout by dissociating our molecular ions state-selectively with methods used in our previous work \cite{ni2014state,zhou2019visible,zhou2020second}. In summary, the state readout consists of the following steps: 
        \begin{enumerate}
            \item  We excite our molecular ions on a bound-to-bound transition with a pulsed laser. This laser is able to resolve electronic, vibrational, and rotational states, but not the hyperfine, parity, and Zeeman manifolds.
            \item We dissociate our state-selectively excited molecular ions with a second pulsed laser to excite them past the dissociation energy into Th$^+$ and F.
            \item Finally, we detect the dissociated Th$^+$ by kicking our ion cloud onto our time-of-flight multi-channel plates, with sufficient temporal resolution to distinguish between dissociated Th$^+$ and residual ThF$^+$.
        \end{enumerate}
    
    \subsection[Microwave spectroscopy on the J=1 to J=2 transition; Hyperfine coupling constant and molecular electric dipole moment]{Microwave spectroscopy on the $J=1$ to $J=2$ transition; Hyperfine coupling constant and molecular electric dipole moment}\label{sec:microwave_spectroscopy}
\begin{figure*}[htb]
            \centering
            \includegraphics[width=1.5\columnwidth]{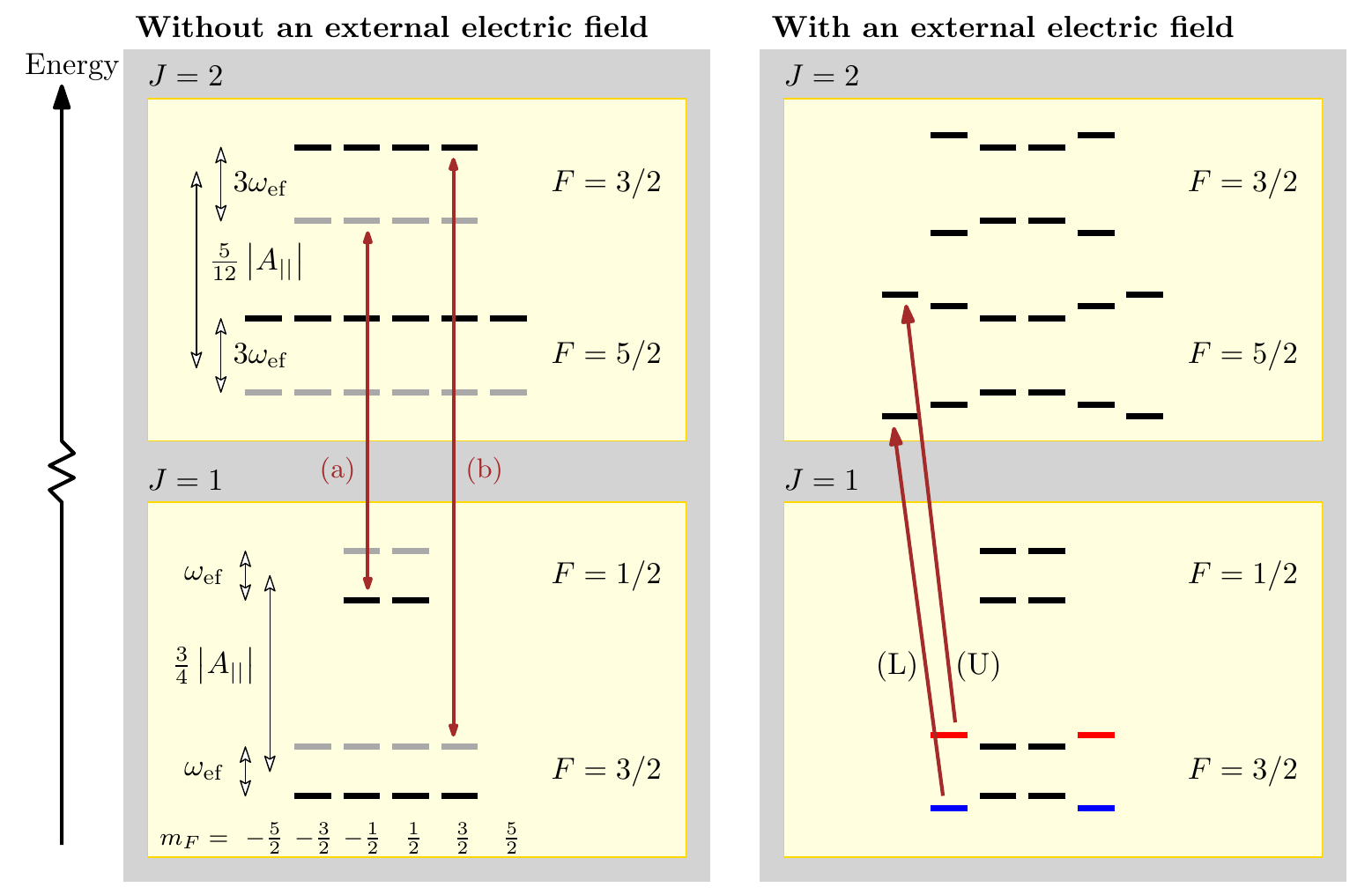}
            \caption{\textbf{Energy landscape (not to scale) of the neighborhood of the eEDM-sensitive state $X\,^3\Delta_1 (v=0$), showing only the first two rotational levels.} In the absence of external electric field, the eigenstates are states of good parity. The energy levels with positive (negative) parity are denoted by black (grey) lines in the energy level diagram on the left. Selection rules for E1 transitions only allow for transitions connecting states of opposite parities, $\Delta F = 0,\pm 1$, and $\Delta m_F = 0,\pm 1$, thus resulting in only six distinct frequencies for resonant transitions connecting the $J=1$ to $J=2$ rotational manifold in the absence of an external magnetic field. For example, the transitions with the highest and lowest frequencies are labelled (a) and (b), respectively, here and also in Figure \ref{fig:microwave0}. In the presence of an external electric field, states of opposite parities mix. The red arrows on the right panel labeled (L) and (U) correspond to the stretched-to-stretched transitions used to determine the Stark shift in our microwave spectroscopy. The Stark shift depends on $m_F$ and $d_\mathrm{mf}$. The eEDM-sensitive states used for the eEDM measurement are the upper (red) and lower (blue) doublets. To reduce clutter in the diagrams, the states drawn do not reflect the true nature of the states in three aspects: (i) the states are drawn with an artificially large $A_{||}$ in comparison to $\omega_\mathrm{ef}$ for well separated hyperfine levels in the diagram; (ii) the Stark shifts in the diagram are portrayed proportionally smaller than those in the actual experiment, where Stark shifts are large enough to allow some states in the lower hyperfine level to be more energetic than the upper; and (iii) $F$ is no longer a good quantum number in the presence of a strong external electric field, except for the most stretched Zeeman states in each rotational manifold. Hence, in the presence of an external electric field, the only selection rule remaining on the $J=1$ to $J=2$ transition is $\Delta m_F = 0,\pm 1$. }
            \label{fig:microwave_energylevel}
        \end{figure*}

        A schematic diagram of the energy levels of the $J=1$ and $J=2$ rotational states in $X\,^3\Delta_1(v=0)$ is shown in Figure \ref{fig:microwave_energylevel}.
        We use state preparation steps shown up to panel (b) in Figure \ref{fig:StatePreparation} to prepare our ions into $X\,^3\Delta_1 (v=0, J=1)$. We then perform microwave spectroscopy on the detailed structure of the $J=1$ to $J=2$ rotational transition. Our observable is the appearance of population in the $J=2$ state, detected by state-selective photodissociation. No external magnetic fields are applied for the microwave spectroscopy experiments in this section.
                
\begin{figure}[htb]
            \centering
            \includegraphics[width=\columnwidth]{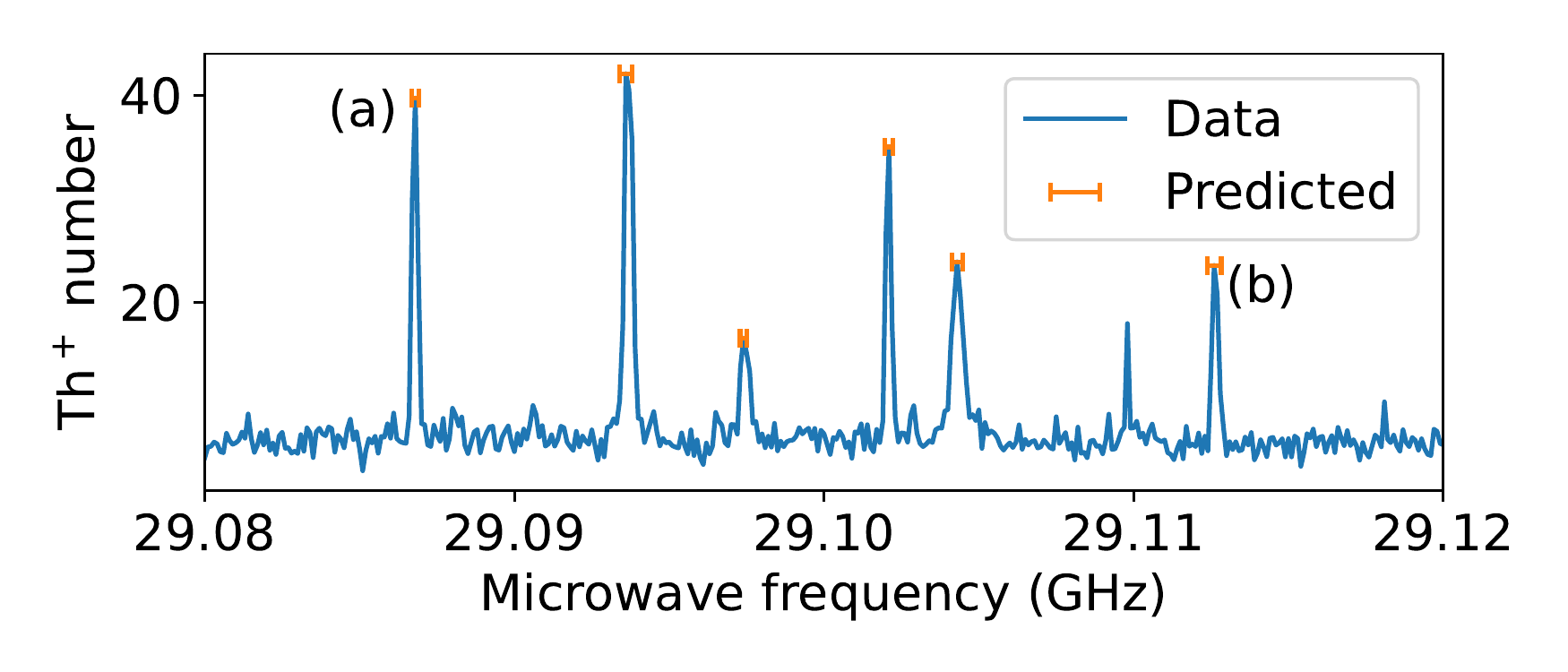}
            \caption{\textbf{Microwave scan at zero net external electric field.} The blue peaks and yellow markers correspond to the actual measurement and simulated position of the peaks, respectively. The unexpected peak at 29.11~GHz is most likely due to a spike in the experimental noise. Error bars from the simulation are propagated from the values shown in Table \ref{tab:sciencestate}. The uncertainty of the Th$^+$ numbers in our data is typically around 5 ions. The lines labelled (a) and (b) are the corresponding transitions labelled in Figure \ref{fig:microwave_energylevel}.}
            \label{fig:microwave0}
        \end{figure}

        At near-zero external electric field, selection rules and energy degeneracies result in just six distinct resonant frequencies (refer to Figures \ref{fig:microwave_energylevel} and \ref{fig:microwave0}), from which we perform a fit to the spectroscopic constants $A_{||}$ (hyperfine coupling constant) and $\omega_\mathrm{ef}$ ($\Omega$-doubling splitting constant). We obtain a $J=2$ to $J=1$ separation of 29.09733(4)~GHz, which is consistent with 29.093(9)~GHz obtained in our previous work \cite{gresh2016broadband}. Figure \ref{fig:microwave_energylevel} illustrates how the spectroscopic constants fit into the energy level structure.
    
\begin{figure}[htb]
            \centering
            \includegraphics[width=\columnwidth]{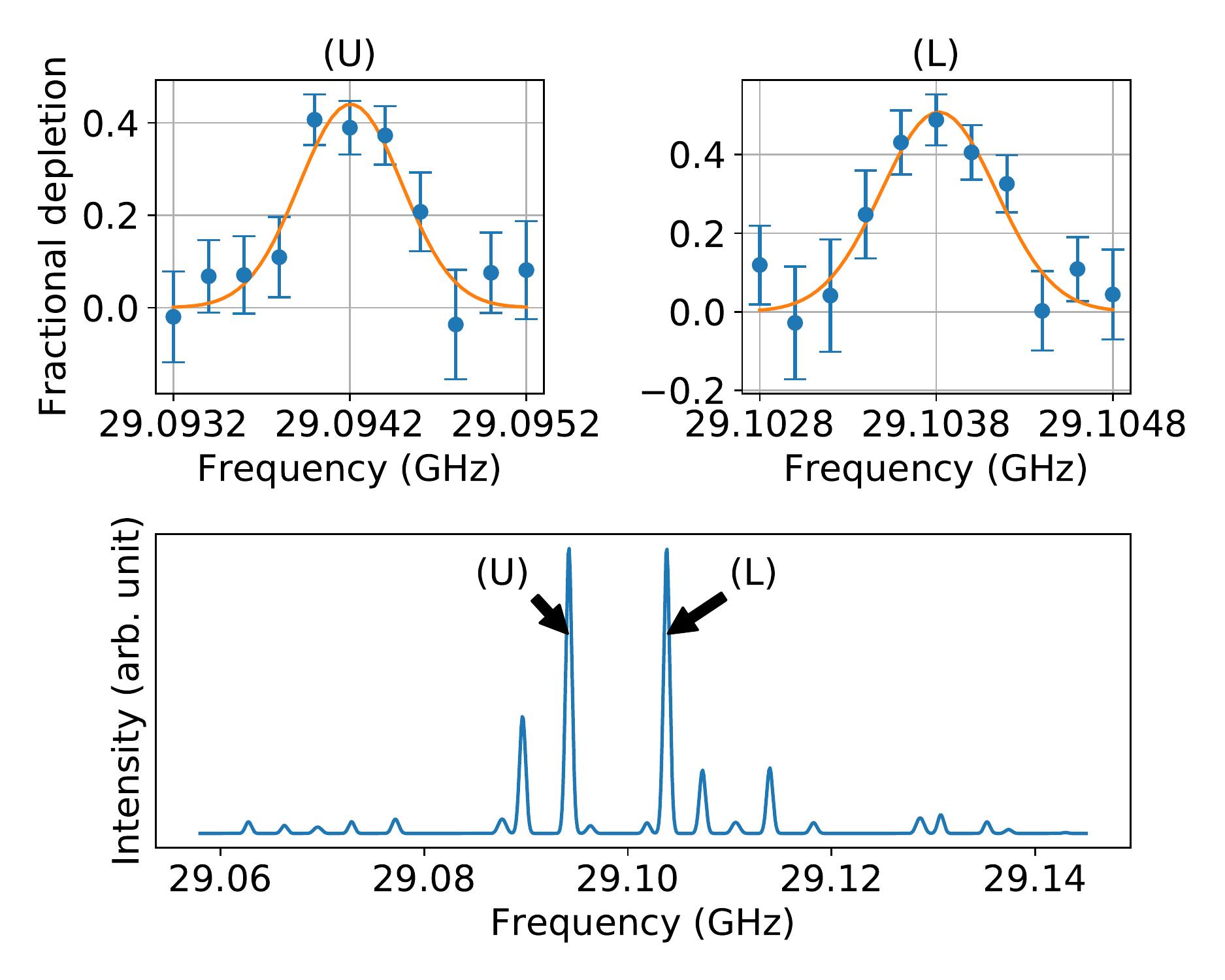}
            \caption{\textbf{Microwave transitions between the $X\,^3\Delta_1 (v=0,J=1,m_F=-3/2)$ and $J=2$ states at an applied electric field strength of 24~V/cm.} The top two plots show the transition lines corresponding to (U) and (L) of Figure \ref{fig:microwave_energylevel}. Error bars show the estimated 1$\sigma$ error in the signal. The bottom plot shows a simulation of the microwave spectrum across a wider frequency range. The intensities are evaluated from Clebsch-Gordan coefficients. In the limit of $d_\mathrm{mf} \mathcal{E} \gg \omega_\mathrm{ef}$, where $\mathcal{E}$ is the applied electric field, the splitting between (U) and (L) is given by $d_\mathrm{mf} \mathcal{E} / 3$. At $\mathcal{E} = 24~\mathrm{V/cm}$, $d_\mathrm{mf} \mathcal{E}$ is roughly 8 times larger than $\omega_\mathrm{ef}$. }
            \label{fig:24V/cm}
        \end{figure}

        By repeating the above experiment with a non-zero external electric field strength, we can see the Stark shifts in the spectral lines (illustrated in Figure \ref{fig:microwave_energylevel}), shifts which depend on $m_F$ and $d_\mathrm{mf}$ (molecular frame electric dipole moment). To enhance signal-to-noise ratio for the extraction of $d_\mathrm{mf}$, we prepare all the ions into $X\,^3\Delta_1 (v=0, J=1, F=3/2, m_F = -3/2)$ with state preparation sequence up to panel (c) in Figure \ref{fig:StatePreparation}. This allows us to suppress all lines coming from the $m_F = \pm1/2$ states to obtain a much cleaner spectrum to extract $d_\mathrm{mf}$. We search for the lines corresponding to the (L) and (U) transitions shown in Figure \ref{fig:microwave_energylevel}. The spectroscopy of these two lines are shown in Figure \ref{fig:24V/cm}.
        These two lines were used for the measurement of $d_\mathrm{mf}$ because of their strong intensities. The (L) and (U) lines will be used to perform doublet depletion [preparation step (d) of Figure \ref{fig:StatePreparation}] in subsequent sections.
        
    \subsection[Ramsey spectroscopy within the J=1 eEDM-sensitive state; Magnetic g-factor]{Ramsey spectroscopy within the $J=1$ eEDM-sensitive state; Magnetic $g$-factor}\label{sec:Ramsey}
        The last spectroscopic constant that we determine is the magnetic $g$-factor for $X\,^3\Delta_1$. We use a rotating electric field to polarize our molecular ions without ejecting them from our ion trap. The rotation micromotion traced out by the ions couples to the applied quadrupole magnetic field gradient to give an averaged net non-zero magnetic field along the instantaneous quantization axis in the frame of the ions, thus resulting in Zeeman shifts of the molecular states. Details of the underlying mechanism can be found in Section 4.11 of Ref.\ \onlinecite{leanhardt2011high}.
        
\begin{figure}[htb]
            \centering
            \includegraphics[width=\columnwidth]{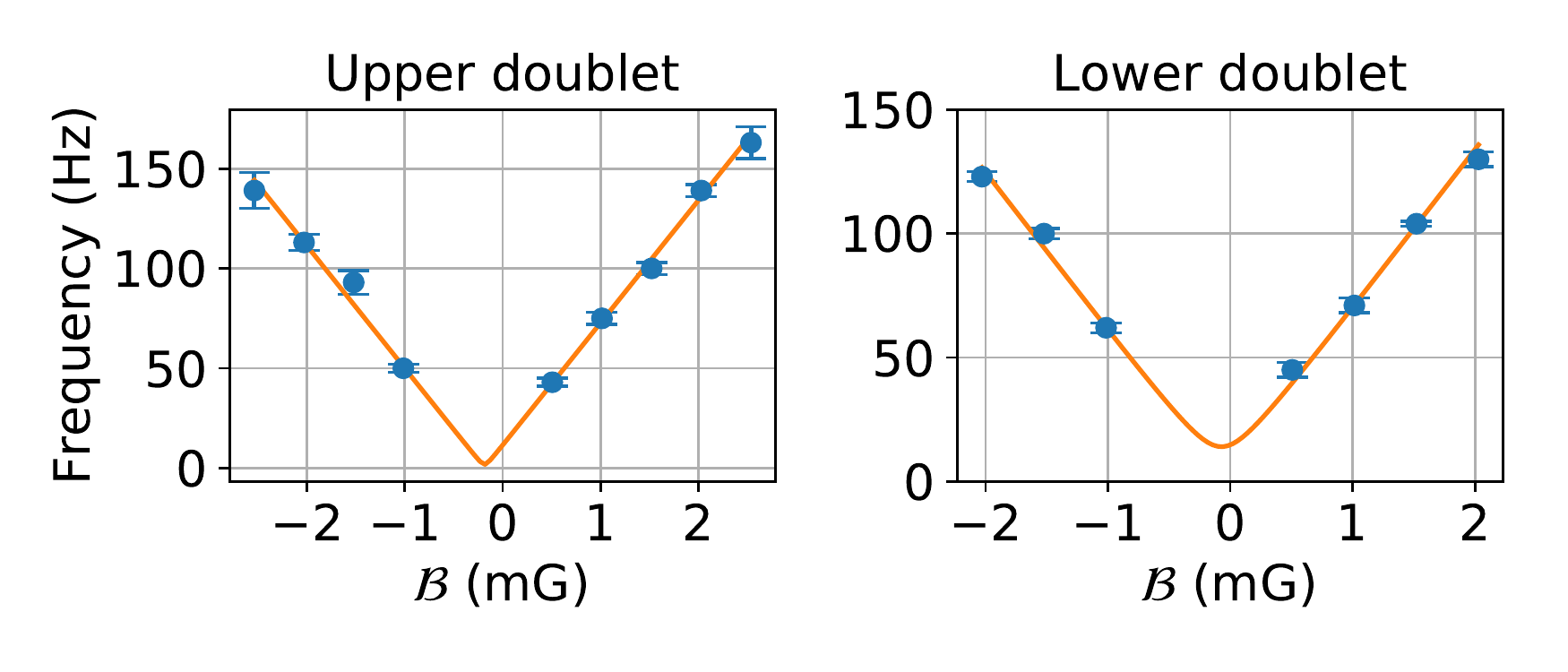}
            \caption{\textbf{Energy differences between the $m_F = \pm 3/2$ states; measurement of the magnetic $g$-factors.} The fits are performed with our model which takes into account the avoided crossing introduced by going into the rotating frame of the molecules. The energy difference at the avoided crossing is fixed by our \textit{ab initio} calculations. The only fit parameters are the asymptotic gradients (corresponding to 3$g_{F=3/2}\mu_\mathrm{B}/h$) and the horizontal offset due to ambient fields. The error bars are 1$\sigma$ error estimates extracted from a non-linear fit to each Ramsey fringe.}
            \label{fig:gfactor}
        \end{figure}

        The rotation frequency is fast compared to trap secular frequencies, but slow compared to typical energy differences between quantum states within the molecular ions. As the ions follow the rotation of the field adiabatically, there is a non-inertial-frame term in the Hamiltonian in the frame of the rotating ions. This non-inertial-frame term introduces rotational coupling between states of $\Delta m_F = \pm 1$, and couples the $m_F = \pm 3/2$ states through a third order process. Restricting ourselves to the Hilbert space involving only the $m_F = \pm 3/2$ states, the good eigenstates of the system are $|m_F = +3/2\rangle \pm |m_F = -3/2\rangle$ at zero external magnetic field, and approaches $|m_F = \pm 3/2\rangle$ asymptotically as the external magnetic field strength increases. Thus, as the strength of the applied magnetic field is swept, the energy difference between the $m_F = \pm 3/2$ states traces out a hyperbola like that shown in Figure \ref{fig:gfactor}, where the vertical offset is due to the avoided crossing introduced by the rotational coupling. 
        
        Following the procedure reported in Ref.\ \onlinecite{loh2013precision}, we map out the the energy differences between the $m_F = \pm 3/2$ states for both the upper and lower doublets by performing Ramsey spectroscopy on $X\,^3\Delta_1$, which is prepared with sequence up to panel (d) in Figure \ref{fig:StatePreparation}. We repeat the experiment at various applied magnetic field strengths. The data and fits are shown in Figure \ref{fig:gfactor}. 
        
        Since we operate at $d_\mathrm{mf} \mathcal{E} \sim A_{||}$, the lower doublet is energetically closer to more $m_F=\pm1/2$ states than the upper doublet. Thus, the lower doublet has a stronger rotational coupling than the upper doublet, resulting in a much larger avoided crossing seen in the plot for the lower doublet than the upper doublet in Figure \ref{fig:gfactor}.
    
        \subsection{Theoretical Calculations of Spectroscopic Constants}\label{sec:theoreticalCalc}
    
        We perform numerical differentiation of coupled-cluster singles and doubles augmented with a non-iterative triples correction [CCSD(T)] \cite{Pople1989} energies to obtain $d_\mathrm{mf}$, $A_{||}$, and $G_{||}$. $G_{||}$ is the response of the electronic energy to magnetic field, as defined in Refs.\ \onlinecite{skripnikov2015theoretical,petrov2017zeeman}. These calculations treat relativistic effects using an exact two-component (X2C) \cite{Dyall1997, XQR} Hamiltonian with atomic mean-field spin-orbit (AMF) integrals \cite{Liu2018}. We use the CFOUR program package \cite{Matthews20CFOUR,cfour} for all the electronic structure calculations presented here.
        We follow the recipe in Ref.\ \onlinecite{Liu2018} for the X2CAMF calculations of $d_\mathrm{mf}$ and $A_{||}$, while we use a unitary transformation scheme \cite{UT} for the calculation of $G_{||}$. Details of the $G_{||}$ calculation will be reported in a separate publication. Calculations of $d_\mathrm{mf}$ and $A_{||}$ use uncontracted ANO-RCC basis sets \cite{ROOS2005, Faegri2001}. We use correlation-consistent polarized core-valence triple- and quadruple-zeta basis sets in the uncontracted form \cite{Th-pwcvtz} to compute $G_{||}$, and we perform basis-set extrapolation to estimate the basis-set-limit value for this property. All CCSD(T) calculations freeze sixty-four core electrons and virtual orbitals higher than 100 hartree. 
            
        Our X2CAMF-CCSD(T) values for $d_\mathrm{mf}$ and $A_{||}$ (Table \ref{tab:sciencestate}) are in fair agreement with the corresponding experimental values. Our computed $d_\mathrm{mf}$ is also in good agreement with calculations from previous work \cite{skripnikov2015theoretical, Denis2015}. Our X2CAMF-CCSD(T)/ANO-RCC-unc value of 6.66~D/{\AA} for the dipole derivative, ${\mathrm{d} (d_\mathrm{mf})}/{\mathrm{d}r}$, predicts a decay lifetime of around 180~ms for the first excited vibronic state, which agrees well with our measurement (Figure \ref{fig:vib_decay}). Our computed $G_{||}$ (Table \ref{tab:sciencestate}) is in reasonable agreement with previous work \cite{skripnikov2015theoretical} and our measured value. It might be of interest to compute the rotational $g$-factor of ThF$^+$, because the rotational $g$-factor has been shown to contribute to about 6\% of the total $g$-factor of a similar molecular species in Ref.\ \onlinecite{ThOgf}.

    \subsection{Summary of Results \& Remarks}\label{sec:summary_spectroscopy}
        Table \ref{tab:sciencestate} shows the measured and calculated spectroscopic structural constants for $X\,^3\Delta_1$.
        \begin{table}[htb]
            \centering
            \begin{tabular}{l l l l}
                \toprule
                Parameters & Exp. & Theory & Previous work \\
                \midrule
                $A_{||}/2\pi$ (MHz) & $-20.1(1)$ & $-21.5$ & - \\
                $\omega_\mathrm{ef}/2\pi$ (MHz) & $5.29(5)$ & - & $5.21(4)$ \cite{gresh2016broadband}\\
                $d_\mathrm{mf}$ (D) & 3.37(9) & 3.46 & 4.03 \cite{Denis2015}, 3.46 \cite{skripnikov2015theoretical} \\
                $|g_{F=3/2}|$ & 0.0149(3) & See main text. & -\\
                $|\delta g_{F=3/2}|$ & 0.0003(3) & - & -\\
                \bottomrule
            \end{tabular}
            \caption{\textbf{Measured spectroscopic structural constants for $X\,^3\Delta_1$.} $A_{||}$, $\omega_\mathrm{ef}$, $d_\mathrm{mf}$, $g_{F=3/2}$, and $\delta g_{F=3/2}$ are the $^{19}$F magnetic hyperfine coupling constant, $\Omega$-doubling splitting constant, molecular frame electric dipole moment of ThF$^+$, the average value of the magnetic $g$-factors of the $F=3/2$ hyperfine level in $X\,^3\Delta_1$ for the upper and lower doublets, and the difference in magnetic $g$-factors between the upper and lower doublets, respectively. Theoretical calculations from this work (details in Section \ref{sec:theoreticalCalc}) and previous work are also shown here for comparison. $A_{||}$ and $\omega_\mathrm{ef}$ have units of $\mathrm{rad}/\mathrm{s}$ and for convenience we divide by $2\pi$ and present our results in millions of cycles per second.}
            \label{tab:sciencestate}
        \end{table}

        Our spectroscopy is not sensitive to the sign of the magnetic $g$-factor shown in Table \ref{tab:sciencestate}. Neglecting the rotational contribution to the $g$-factor, and converting $|g_{F=3/2}|$ into $G_{||}$, we get $-0.042(2)$ if $g_{F=3/2} > 0$ and $0.048(2)$ otherwise. The latter is not far away from the theoretical predictions of $G_{||} = 0.034$ \cite{skripnikov2015theoretical} and 0.035 calculated in this work. We do not have a systematic estimate for the error in the theoretical value of $G_{||}$, therefore there remains some ambiguity in sign of the $g$-factor.
        
        The spectroscopic constants measured in ThF$^+$ are similar to those in HfF$^+$ \cite{loh2013precision}. This means that we will be operating in a familiar experimental parameter space. Therefore, the eEDM experimental complexity will not increase with the planned molecule upgrade from HfF$^+$ to ThF$^+$.

\section[Blackbody Radiation Excitation and T1 Relaxation Time]{Blackbody Radiation Excitation and $T_1$ Relaxation Time}\label{sec:BBR}
    $X\,^3\Delta_1$ has been shown to be the ground state of ThF$^+$ \cite{gresh2016broadband}. Hence the coherence time of $X\,^3\Delta_1$ is not subjected to spontaneous decay. However, stray photons, e.g.\ blackbody radiation, can excite ThF$^+$ from $X\,^3\Delta_1$, and they subsequently either decay into other long-lived states where they no longer contribute to the measurement statistics, or back into $X\,^3\Delta_1$ with corresponding delayed decay in coherent spectroscopy contrast. 
    
\begin{figure}[htb]
        \centering
        \includegraphics[width=0.9\columnwidth]{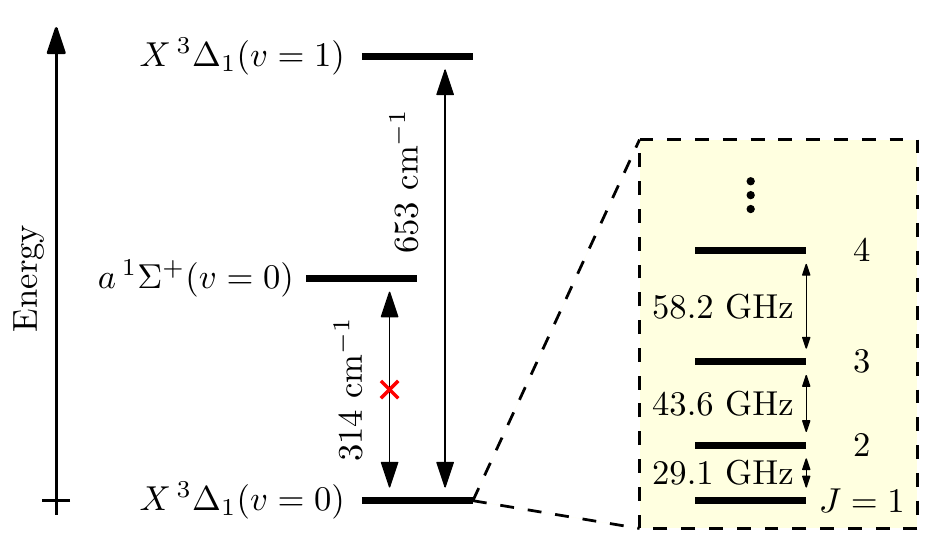}
        \caption{\textbf{Energy level diagram of the lowest few vibronic states in ThF$^+$.} The transition between $X\,^3\Delta_1$ and $a\,^1\Sigma^+$ is forbidden by selection rules. A zoomed-in view of the rotational states within the ground vibronic state is shown in the dashed box. $1~\mathrm{cm}^{-1} \approx 30~\mathrm{GHz}$.}
        \label{fig:EnergyLevel_BBR}
    \end{figure}

    An energy level diagram of the lowest few vibronic states in ThF$^+$ is shown in Figure \ref{fig:EnergyLevel_BBR}.
    At room temperature, the energy of a blackbody radiation photon at peak intensity is on the order of the vibrational spacing in $X\,^3\Delta_1$. The dominant blackbody radiation excitation channel from $X\,^3\Delta_1(v=0,J=1)$ is through $X\,^3\Delta_1(v=1)$, because (i) the lowest vibrational states are highly harmonic such that only $\Delta v = \pm 1$ transitions are allowed, (ii) rotational spacing is too small for appreciable transition rates between rotational states, and (iii) selection rules forbid transitions between $X\,^3\Delta_1$ and $a^1\Sigma^+$. 
    
    The blackbody radiation excitation rate from $X\,^3\Delta_1(v=0)$ through $X\,^3\Delta_1(v=1)$ is related to the spontaneous decay lifetime of the $v=1$ states. This relation comes through the transition dipole moment between the $v=0$ and $v=1$ states. Hence, a measurement of the lifetime of the $v=1$ state will allow us to predict the blackbody radiation excitation rate from $X\,^3\Delta_1$.
    
\begin{figure}[htb]
        \centering
        \includegraphics[width=\columnwidth]{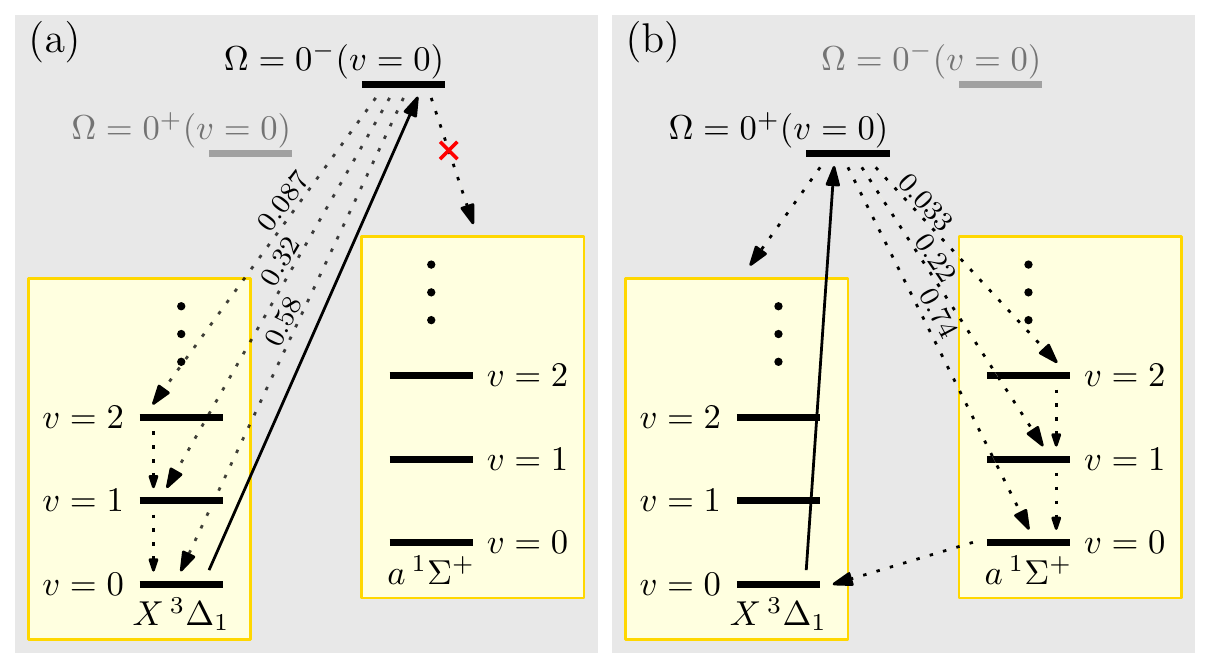}
        \caption{\textbf{State preparation for measurement of excited state lifetime.} (a) For measurement of $X\,^3\Delta_1(v=1)$ state lifetime. (b) For measurement of $a\,^1\Sigma^+(v=0)$ state lifetime. Solid line indicates laser used for optical pumping. Dotted lines indicate spontaneous emission. Numbers attached to dotted lines indicate Franck-Condon factors determined from our previous spectroscopy work \cite{gresh2016broadband}.}
        \label{fig:Vibrational_lifetime_state_prep}
    \end{figure}

\begin{figure}[htb]
        \centering
        \includegraphics[width=\columnwidth]{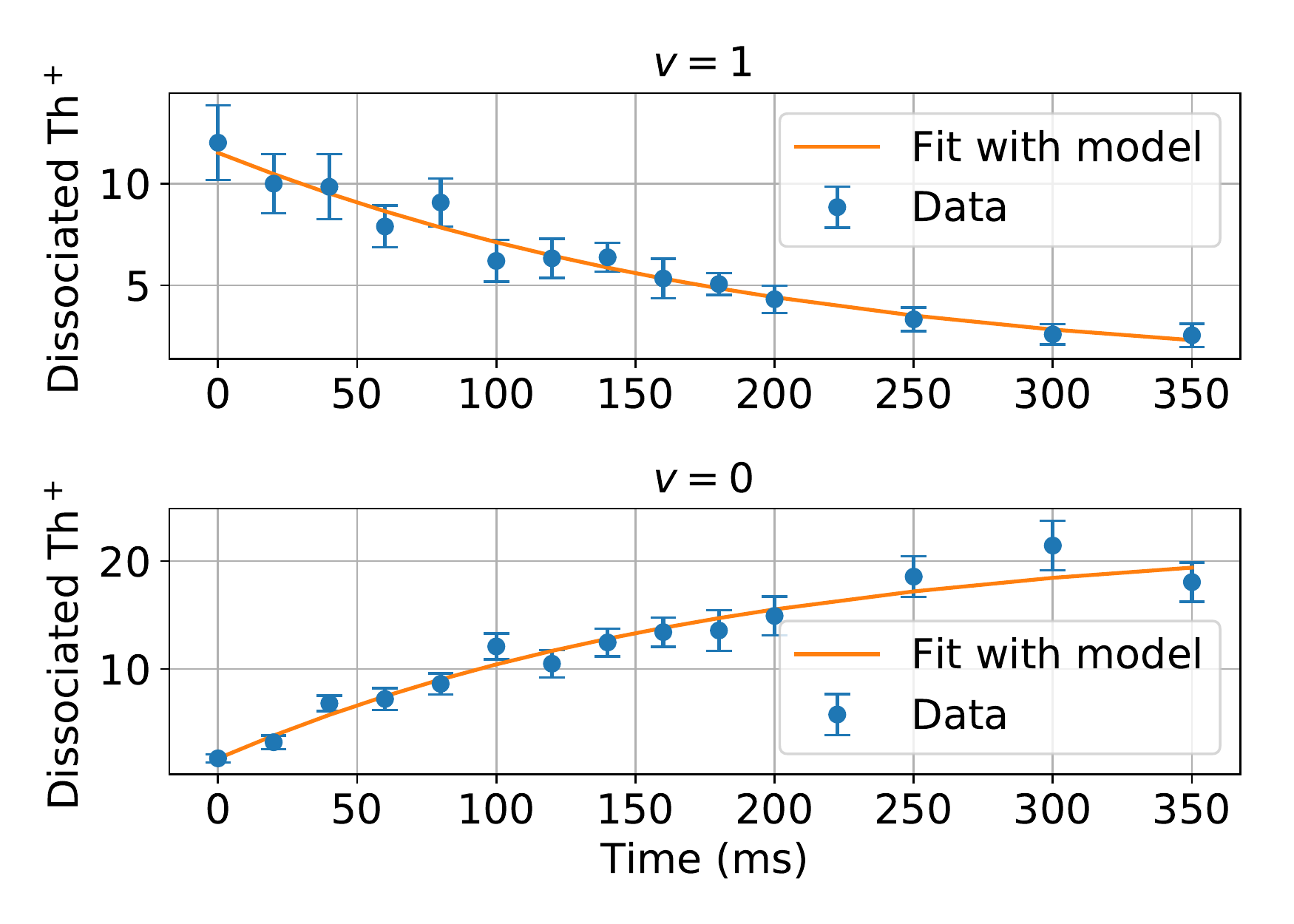}
        \caption{\textbf{Vibrational population decay from $v=1$ to $v=0$ of the ground electronic state in ThF$^+$.} All the population was pumped from the ground vibrational manifold into the excited vibrational manifold through the $\Omega=0^-$ state \cite{gresh2016broadband}. Population in each vibrational manifold is read out with photodissociation. The observed vibrational decay lifetime agrees with our model, which is used to predict the blackbody radiation excitation lifetime from $X\,^3\Delta_1$. The total dissociated Th$^+$ numbers do not appear to be conserved in these two plots because the dissociation efficiencies for these two vibrational manifolds are not the same.}
        \label{fig:vib_decay}
    \end{figure}

    To measure the spontaneous decay lifetime of the $v=1$ state, we first prepare all our ions in the ground vibronic state $X\,^3\Delta_1(v=0)$ selectively with resonance-enhanced--multi-photon ionization \cite{zhou2019visible}, and optically excite all our ions to the $\Omega=0^-$ excited state \cite{gresh2016broadband}, allowing for the ions to decay back into the $X\,^3\Delta_1(v>0)$ states, as illustrated in Figure \ref{fig:Vibrational_lifetime_state_prep}(a).
    We note that the branching ratio from the $\Omega=0^-$ state to the $X\,^3\Delta_1$ manifold is very close to unity. Off-diagonal Franck-Condon factors allow the $v\geq1$ manifolds to be populated through optical pumping. The whole optical pumping process takes about 100~ms. We then allow the ions to decay from the excited vibrational states to lower ones, and read out the population in each vibrational state with resonance-enhanced--multi-photon dissociation \cite{zhou2019visible,ni2014state} much like how we have described in Section \ref{sec:setup}. We dissociate through the R(1) line for both the $v=0$ and $v=1$ manifolds. Our result is shown in Figure \ref{fig:vib_decay}.
    
    To extract the spontaneous decay lifetime of the $v=1$ state from Figure \ref{fig:vib_decay}, we employ the following model:
    \begin{enumerate}
        \item We allow for non-zero initial population in the $v=1$ and $v=2$ states. The $v=1$ and $v=2$ states are populated by decay from the $\Omega=0^-(v'=0)$ state, and the Franck-Condon factors are such that we may approximate the initial populations of $v \geq 3$ as zero.
        \item The decay rates from each vibrational manifold are governed by their respective Einstein's $A$ coefficients. We also assume that only $\Delta v=\pm 1$ transitions are allowed.
        \item Fit parameters include (i) absolute scaling to account for different detection efficiencies for the $v=0$ and $v=1$ manifolds; (ii) number of background ions; (iii) ratio of initial populations in the $v=1$ and $v=2$ manifolds; and (iv) ${\mathrm{d} (d_\mathrm{mf})}/{\mathrm{d}r}$, with this being held the same across the fits for $v=0$ and $v=1$.
    \end{enumerate}
    Note that we can group (i) the effect of imperfect optical pumping from the $v=0$ manifold, which results in an initial non-zero population in the $v=0$ manifold, and (ii) background ions detected for the $v=0$ manifold together into a single $v=0$ ``background ion'' fit parameter in the fitting model. With the above model, we obtain $\mathrm{d}(d_\mathrm{mf})/\mathrm{d}r = 7(2)~\mathrm{D}/\mathrm{\AA}$, in good agreement with our calculated value from Section \ref{sec:theoreticalCalc}. This value corresponds to spontaneous decay lifetimes of 0.16(11)~s and 0.08(6)~s for $v=1 \rightarrow v=0$ and $v=2 \rightarrow v=1$, respectively.
    
    Using the value of the molecular dipole, we predict the effective lifetime ($T_1$ relaxation time) of $X\,^3\Delta_1$ to be about 3~s at room temperature (300~K), which is limited by blackbody radiation excitation from $X\,^3\Delta_1$ to the first excited vibronic state $X\,^3\Delta_1(v=1)$. 
    
    We can suppress blackbody radiation excitation from $X\,^3\Delta_1(v=0)$ by introducing cryogenics to lower the temperature of the setup. The effective lifetime of $X\,^3\Delta_1(v=0)$ increases drastically with a decrease in temperature (Table \ref{tab:BBR_vib}), and we anticipate establishing a blackbody environment at 180~K, which will be a workable balance between technical convenience and sufficiently long interrogation times.
    \begin{table}
        \centering
        \begin{tabular}{l c c c c c c}
            \toprule
             & 77~K & 120~K & 150~K & 180~K & 200~K & 300~K \\
            \midrule
            To $X\,^3\Delta_1(v=1)$ & 32000 & 400 & 84 & 29 & 17 & 3.5 \\
            To $a\,^1\Sigma^+$ & 2200 & 270 & 130 & 80 & 64 & 33 \\
            \midrule
            Combined & 2100 & 160 & 51 & 21 & 13 & 3.2 \\
            \bottomrule
        \end{tabular}
        \caption{\textbf{Prediction of blackbody radiation excitation lifetime (in seconds) out of the $v=0$ vibrational level in $X\,^3\Delta_1$, in a radiative environment of the indicated temperature.}}
        \label{tab:BBR_vib}
    \end{table}
    
    We calculate blackbody radiation excitation across rotational levels to occur at a time scale of $10^3$ seconds and above, even at 300~K. Since we plan to measure the eEDM with an interrogation time of about 20~s, which is only about 10 times longer than our current experiment using HfF$^+$, the effects of rotational blackbody radiation excitation are small and shall be neglected for the rest of the discussion. 
    
    Usual selection rules forbid transition between $X\,^3\Delta_1$ and $a\,^1\Sigma^+$, but our molecular ion falls under Hund's case (c), so these states contain slight admixtures of states of other character. Hence, a transition between $X\,^3\Delta_1$ and $a\,^1\Sigma^+$ is not entirely forbidden. By using an $\Omega=0^+$ state \cite{gresh2016broadband} that couples both to $X\,^3\Delta_1$ and $a\,^1\Sigma^+$ for optical pumping, we populate the $a\,^1\Sigma^+ (v=0)$ manifold through a process similar to the experiment for measuring the $X\,^3\Delta_1 (v=1)$ spontaneous decay lifetime [refer to Figure \ref{fig:Vibrational_lifetime_state_prep}(b)]. We observe the spontaneous decay lifetime from $a\,^1\Sigma^+ (v=0)$ back to $X\,^3\Delta_1 (v=0)$ to be about 6~s, which is about 40 times longer than from $X\,^3\Delta_1 (v=1)$. We calculate blackbody radiation excitation from $X\,^3\Delta_1$ to $a\,^1\Sigma^+$ to be in excess of $30$ seconds (Table \ref{tab:BBR_vib}).
    
    The net effect of blackbody radiation excitations to $a\,^1\Sigma^+$ and $X\,^3\Delta_1(v=1)$ at 180~K gives an expected lifetime of $X\,^3\Delta_1$ to be about 20~s.

\section{Conclusion \& Outlook}
    We perform spectroscopy on $X\,^3\Delta_1$ to extract its spectroscopic constants.
    We also measure the state lifetime of the first excited vibrational state and show that it is consistent with our \textit{ab initio} calculations. Given this assurance, we predict that an eEDM experiment with ThF$^+$ in a 180~K environment is sufficient to achieve a 20~s coherence time, a ten-times improvement over our ongoing eEDM experiment with HfF$^+$. 
    
    The stage is set for performing an eEDM measurement with ThF$^+$. We expect a significant improvement in statistical sensitivity in the measurement over the HfF$^+$ system. Ongoing work includes testing out the modest cryogenic system at 180~K to suppress blackbody radiation, and multiplexing the experiment with a conveyor belt of ion traps to increase count rates.
    
\section*{Acknowledgements}
    This work is supported by Moore Foundation, Sloan Foundation, NSF PFC 1734006, NIST, and Marsico Research Chair.
    
    Computational work at Johns Hopkins University is supported by National Science Foundation under grant number PHY-2011794.

\bibliography{biblio}

\begin{thebibliography}{46}%
\makeatletter
\providecommand \@ifxundefined [1]{%
 \@ifx{#1\undefined}
}%
\providecommand \@ifnum [1]{%
 \ifnum #1\expandafter \@firstoftwo
 \else \expandafter \@secondoftwo
 \fi
}%
\providecommand \@ifx [1]{%
 \ifx #1\expandafter \@firstoftwo
 \else \expandafter \@secondoftwo
 \fi
}%
\providecommand \natexlab [1]{#1}%
\providecommand \enquote  [1]{``#1''}%
\providecommand \bibnamefont  [1]{#1}%
\providecommand \bibfnamefont [1]{#1}%
\providecommand \citenamefont [1]{#1}%
\providecommand \href@noop [0]{\@secondoftwo}%
\providecommand \href [0]{\begingroup \@sanitize@url \@href}%
\providecommand \@href[1]{\@@startlink{#1}\@@href}%
\providecommand \@@href[1]{\endgroup#1\@@endlink}%
\providecommand \@sanitize@url [0]{\catcode `\\12\catcode `\$12\catcode
  `\&12\catcode `\#12\catcode `\^12\catcode `\_12\catcode `\%12\relax}%
\providecommand \@@startlink[1]{}%
\providecommand \@@endlink[0]{}%
\providecommand \url  [0]{\begingroup\@sanitize@url \@url }%
\providecommand \@url [1]{\endgroup\@href {#1}{\urlprefix }}%
\providecommand \urlprefix  [0]{URL }%
\providecommand \Eprint [0]{\href }%
\providecommand \doibase [0]{https://doi.org/}%
\providecommand \selectlanguage [0]{\@gobble}%
\providecommand \bibinfo  [0]{\@secondoftwo}%
\providecommand \bibfield  [0]{\@secondoftwo}%
\providecommand \translation [1]{[#1]}%
\providecommand \BibitemOpen [0]{}%
\providecommand \bibitemStop [0]{}%
\providecommand \bibitemNoStop [0]{.\EOS\space}%
\providecommand \EOS [0]{\spacefactor3000\relax}%
\providecommand \BibitemShut  [1]{\csname bibitem#1\endcsname}%
\let\auto@bib@innerbib\@empty
\bibitem [{\citenamefont {Hinds}(1997)}]{hinds1997testing}%
  \BibitemOpen
  \bibfield  {author} {\bibinfo {author} {\bibfnamefont {E.~A.}\ \bibnamefont
  {Hinds}},\ }\bibfield  {title} {\bibinfo {title} {Testing time reversal
  symmetry using molecules},\ }\href@noop {} {\bibfield  {journal} {\bibinfo
  {journal} {Physica Scripta}\ }\textbf {\bibinfo {volume} {1997}},\ \bibinfo
  {pages} {34} (\bibinfo {year} {1997})}\BibitemShut {NoStop}%
\bibitem [{\citenamefont {Khriplovich}\ and\ \citenamefont
  {Lamoreaux}(2012)}]{khriplovich2012cp}%
  \BibitemOpen
  \bibfield  {author} {\bibinfo {author} {\bibfnamefont {I.~B.}\ \bibnamefont
  {Khriplovich}}\ and\ \bibinfo {author} {\bibfnamefont {S.~K.}\ \bibnamefont
  {Lamoreaux}},\ }\href@noop {} {\emph {\bibinfo {title} {{CP} violation
  without strangeness: electric dipole moments of particles, atoms, and
  molecules}}}\ (\bibinfo  {publisher} {Springer Science \& Business Media},\
  \bibinfo {year} {2012})\BibitemShut {NoStop}%
\bibitem [{\citenamefont {Chupp}\ and\ \citenamefont
  {Ramsey-Musolf}(2015)}]{Chupp2015}%
  \BibitemOpen
  \bibfield  {author} {\bibinfo {author} {\bibfnamefont {T.}~\bibnamefont
  {Chupp}}\ and\ \bibinfo {author} {\bibfnamefont {M.}~\bibnamefont
  {Ramsey-Musolf}},\ }\bibfield  {title} {\bibinfo {title} {Electric dipole
  moments: A global analysis},\ }\href@noop {} {\bibfield  {journal} {\bibinfo
  {journal} {Physical Review C - Nuclear Physics}\ }\textbf {\bibinfo {volume}
  {91}},\ \bibinfo {pages} {1} (\bibinfo {year} {2015})}\BibitemShut {NoStop}%
\bibitem [{\citenamefont {Stadnik}\ \emph {et~al.}()\citenamefont {Stadnik},
  \citenamefont {Dzuba},\ and\ \citenamefont {Flambaum}}]{Stadnik2018}%
  \BibitemOpen
  \bibfield  {author} {\bibinfo {author} {\bibfnamefont {Y.~V.}\ \bibnamefont
  {Stadnik}}, \bibinfo {author} {\bibfnamefont {V.~A.}\ \bibnamefont {Dzuba}},\
  and\ \bibinfo {author} {\bibfnamefont {V.~V.}\ \bibnamefont {Flambaum}},\
  }\bibfield  {title} {\bibinfo {title} {Improved limits on
  axionlike-particle-mediated {P}, {T}-violating interactions between electrons
  and nucleons from electric dipole moments of atoms and molecules},\
  }\href@noop {} {\bibinfo  {journal} {Physical Review Letters}\ ,\ \bibinfo
  {pages} {13202}}\BibitemShut {NoStop}%
\bibitem [{\citenamefont {Cesarotti}\ \emph {et~al.}(2019)\citenamefont
  {Cesarotti}, \citenamefont {Lu}, \citenamefont {Nakai}, \citenamefont
  {Parikh},\ and\ \citenamefont {Reece}}]{cesarotti2019interpreting}%
  \BibitemOpen
\bibfield  {journal} {  }\bibfield  {author} {\bibinfo {author} {\bibfnamefont
  {C.}~\bibnamefont {Cesarotti}}, \bibinfo {author} {\bibfnamefont
  {Q.}~\bibnamefont {Lu}}, \bibinfo {author} {\bibfnamefont {Y.}~\bibnamefont
  {Nakai}}, \bibinfo {author} {\bibfnamefont {A.}~\bibnamefont {Parikh}},\ and\
  \bibinfo {author} {\bibfnamefont {M.}~\bibnamefont {Reece}},\ }\bibfield
  {title} {\bibinfo {title} {Interpreting the electron {EDM} constraint},\
  }\href@noop {} {\bibfield  {journal} {\bibinfo  {journal} {Journal of High
  Energy Physics}\ }\textbf {\bibinfo {volume} {2019}},\ \bibinfo {pages} {1}
  (\bibinfo {year} {2019})}\BibitemShut {NoStop}%
\bibitem [{\citenamefont {Sarkar}(1996)}]{sarkar1996big}%
  \BibitemOpen
  \bibfield  {author} {\bibinfo {author} {\bibfnamefont {S.}~\bibnamefont
  {Sarkar}},\ }\bibfield  {title} {\bibinfo {title} {Big bang nucleosynthesis
  and physics beyond the standard model},\ }\href@noop {} {\bibfield  {journal}
  {\bibinfo  {journal} {Reports on Progress in Physics}\ }\textbf {\bibinfo
  {volume} {59}},\ \bibinfo {pages} {1493} (\bibinfo {year}
  {1996})}\BibitemShut {NoStop}%
\bibitem [{\citenamefont {Ellis}(2007)}]{ellis2007beyond}%
  \BibitemOpen
  \bibfield  {author} {\bibinfo {author} {\bibfnamefont {J.}~\bibnamefont
  {Ellis}},\ }\bibfield  {title} {\bibinfo {title} {Beyond the standard model
  with the {LHC}},\ }\href@noop {} {\bibfield  {journal} {\bibinfo  {journal}
  {Nature}\ }\textbf {\bibinfo {volume} {448}},\ \bibinfo {pages} {297}
  (\bibinfo {year} {2007})}\BibitemShut {NoStop}%
\bibitem [{\citenamefont {Pospelov}\ and\ \citenamefont
  {Khriplovich}(1991)}]{pospelov1991electric}%
  \BibitemOpen
  \bibfield  {author} {\bibinfo {author} {\bibfnamefont {M.~E.}\ \bibnamefont
  {Pospelov}}\ and\ \bibinfo {author} {\bibfnamefont {I.}~\bibnamefont
  {Khriplovich}},\ }\bibfield  {title} {\bibinfo {title} {Electric dipole
  moment of the {W} boson and the electron in the {Kobayashi-Maskawa} model},\
  }\href@noop {} {\bibfield  {journal} {\bibinfo  {journal} {Yadernaya Fizika}\
  }\textbf {\bibinfo {volume} {53}},\ \bibinfo {pages} {1030} (\bibinfo {year}
  {1991})}\BibitemShut {NoStop}%
\bibitem [{\citenamefont {Barr}\ and\ \citenamefont
  {Zee}(1990)}]{barr1990electric}%
  \BibitemOpen
  \bibfield  {author} {\bibinfo {author} {\bibfnamefont {S.~M.}\ \bibnamefont
  {Barr}}\ and\ \bibinfo {author} {\bibfnamefont {A.}~\bibnamefont {Zee}},\
  }\bibfield  {title} {\bibinfo {title} {Electric dipole moment of the electron
  and of the neutron},\ }\href@noop {} {\bibfield  {journal} {\bibinfo
  {journal} {Physical Review Letters}\ }\textbf {\bibinfo {volume} {65}},\
  \bibinfo {pages} {21} (\bibinfo {year} {1990})}\BibitemShut {NoStop}%
\bibitem [{\citenamefont {Bernreuther}\ and\ \citenamefont
  {Suzuki}(1991)}]{bernreuther1991electric}%
  \BibitemOpen
  \bibfield  {author} {\bibinfo {author} {\bibfnamefont {W.}~\bibnamefont
  {Bernreuther}}\ and\ \bibinfo {author} {\bibfnamefont {M.}~\bibnamefont
  {Suzuki}},\ }\bibfield  {title} {\bibinfo {title} {The electric dipole moment
  of the electron},\ }\href@noop {} {\bibfield  {journal} {\bibinfo  {journal}
  {Reviews of Modern Physics}\ }\textbf {\bibinfo {volume} {63}},\ \bibinfo
  {pages} {313} (\bibinfo {year} {1991})}\BibitemShut {NoStop}%
\bibitem [{\citenamefont {Barr}(1993)}]{barr1993review}%
  \BibitemOpen
  \bibfield  {author} {\bibinfo {author} {\bibfnamefont {S.~M.}\ \bibnamefont
  {Barr}},\ }\bibfield  {title} {\bibinfo {title} {A review of {CP} violation
  in atoms},\ }\href@noop {} {\bibfield  {journal} {\bibinfo  {journal}
  {International Journal of Modern Physics A}\ }\textbf {\bibinfo {volume}
  {8}},\ \bibinfo {pages} {209} (\bibinfo {year} {1993})}\BibitemShut {NoStop}%
\bibitem [{\citenamefont {Pospelov}\ and\ \citenamefont
  {Ritz}(2005)}]{pospelov2005electric}%
  \BibitemOpen
  \bibfield  {author} {\bibinfo {author} {\bibfnamefont {M.}~\bibnamefont
  {Pospelov}}\ and\ \bibinfo {author} {\bibfnamefont {A.}~\bibnamefont
  {Ritz}},\ }\bibfield  {title} {\bibinfo {title} {Electric dipole moments as
  probes of new physics},\ }\href@noop {} {\bibfield  {journal} {\bibinfo
  {journal} {Annals of physics}\ }\textbf {\bibinfo {volume} {318}},\ \bibinfo
  {pages} {119} (\bibinfo {year} {2005})}\BibitemShut {NoStop}%
\bibitem [{\citenamefont {Commins}(1999)}]{commins1999electric}%
  \BibitemOpen
  \bibfield  {author} {\bibinfo {author} {\bibfnamefont {E.~D.}\ \bibnamefont
  {Commins}},\ }\bibfield  {title} {\bibinfo {title} {Electric dipole moments
  of leptons},\ }in\ \href@noop {} {\emph {\bibinfo {booktitle} {Advances in
  Atomic, Molecular, and Optical Physics}}},\ Vol.~\bibinfo {volume} {40}\
  (\bibinfo  {publisher} {Elsevier},\ \bibinfo {year} {1999})\ pp.\ \bibinfo
  {pages} {1--55}\BibitemShut {NoStop}%
\bibitem [{\citenamefont {Cairncross}\ \emph {et~al.}(2017)\citenamefont
  {Cairncross}, \citenamefont {Gresh}, \citenamefont {Grau}, \citenamefont
  {Cossel}, \citenamefont {Roussy}, \citenamefont {Ni}, \citenamefont {Zhou},
  \citenamefont {Ye},\ and\ \citenamefont {Cornell}}]{cairncross2017precision}%
  \BibitemOpen
  \bibfield  {author} {\bibinfo {author} {\bibfnamefont {W.~B.}\ \bibnamefont
  {Cairncross}}, \bibinfo {author} {\bibfnamefont {D.~N.}\ \bibnamefont
  {Gresh}}, \bibinfo {author} {\bibfnamefont {M.}~\bibnamefont {Grau}},
  \bibinfo {author} {\bibfnamefont {K.~C.}\ \bibnamefont {Cossel}}, \bibinfo
  {author} {\bibfnamefont {T.~S.}\ \bibnamefont {Roussy}}, \bibinfo {author}
  {\bibfnamefont {Y.}~\bibnamefont {Ni}}, \bibinfo {author} {\bibfnamefont
  {Y.}~\bibnamefont {Zhou}}, \bibinfo {author} {\bibfnamefont {J.}~\bibnamefont
  {Ye}},\ and\ \bibinfo {author} {\bibfnamefont {E.~A.}\ \bibnamefont
  {Cornell}},\ }\bibfield  {title} {\bibinfo {title} {Precision measurement of
  the {electron's} electric dipole moment using trapped molecular ions},\
  }\href@noop {} {\bibfield  {journal} {\bibinfo  {journal} {Physical Review
  Letters}\ }\textbf {\bibinfo {volume} {119}},\ \bibinfo {pages} {153001}
  (\bibinfo {year} {2017})}\BibitemShut {NoStop}%
\bibitem [{\citenamefont {Andreev}\ \emph {et~al.}(2018)\citenamefont
  {Andreev}, \citenamefont {Ang}, \citenamefont {DeMille}, \citenamefont
  {Doyle}, \citenamefont {Gabrielse}, \citenamefont {Haefner}, \citenamefont
  {Hutzler}, \citenamefont {Lasner}, \citenamefont {Meisenhelder},
  \citenamefont {O'Leary}, \citenamefont {Panda}, \citenamefont {West},
  \citenamefont {West},\ and\ \citenamefont {Wu}}]{acme2018improved}%
  \BibitemOpen
  \bibfield  {author} {\bibinfo {author} {\bibfnamefont {V.}~\bibnamefont
  {Andreev}}, \bibinfo {author} {\bibfnamefont {D.~G.}\ \bibnamefont {Ang}},
  \bibinfo {author} {\bibfnamefont {D.}~\bibnamefont {DeMille}}, \bibinfo
  {author} {\bibfnamefont {J.~M.}\ \bibnamefont {Doyle}}, \bibinfo {author}
  {\bibfnamefont {G.}~\bibnamefont {Gabrielse}}, \bibinfo {author}
  {\bibfnamefont {J.}~\bibnamefont {Haefner}}, \bibinfo {author} {\bibfnamefont
  {N.~R.}\ \bibnamefont {Hutzler}}, \bibinfo {author} {\bibfnamefont
  {Z.}~\bibnamefont {Lasner}}, \bibinfo {author} {\bibfnamefont
  {C.}~\bibnamefont {Meisenhelder}}, \bibinfo {author} {\bibfnamefont {B.~R.}\
  \bibnamefont {O'Leary}}, \bibinfo {author} {\bibfnamefont {C.~D.}\
  \bibnamefont {Panda}}, \bibinfo {author} {\bibfnamefont {A.~D.}\ \bibnamefont
  {West}}, \bibinfo {author} {\bibfnamefont {E.~P.}\ \bibnamefont {West}},\
  and\ \bibinfo {author} {\bibfnamefont {X.}~\bibnamefont {Wu}},\ }\bibfield
  {title} {\bibinfo {title} {Improved limit on the electric dipole moment of
  the electron},\ }\href@noop {} {\bibfield  {journal} {\bibinfo  {journal}
  {Nature}\ }\textbf {\bibinfo {volume} {562}},\ \bibinfo {pages} {355}
  (\bibinfo {year} {2018})}\BibitemShut {NoStop}%
\bibitem [{\citenamefont {Hudson}\ \emph {et~al.}(2011)\citenamefont {Hudson},
  \citenamefont {Kara}, \citenamefont {Smallman}, \citenamefont {Sauer},
  \citenamefont {Tarbutt},\ and\ \citenamefont {Hinds}}]{hudson2011improved}%
  \BibitemOpen
  \bibfield  {author} {\bibinfo {author} {\bibfnamefont {J.~J.}\ \bibnamefont
  {Hudson}}, \bibinfo {author} {\bibfnamefont {D.~M.}\ \bibnamefont {Kara}},
  \bibinfo {author} {\bibfnamefont {I.~J.}\ \bibnamefont {Smallman}}, \bibinfo
  {author} {\bibfnamefont {B.~E.}\ \bibnamefont {Sauer}}, \bibinfo {author}
  {\bibfnamefont {M.~R.}\ \bibnamefont {Tarbutt}},\ and\ \bibinfo {author}
  {\bibfnamefont {E.~A.}\ \bibnamefont {Hinds}},\ }\bibfield  {title} {\bibinfo
  {title} {Improved measurement of the shape of the electron},\ }\href@noop {}
  {\bibfield  {journal} {\bibinfo  {journal} {Nature}\ }\textbf {\bibinfo
  {volume} {473}},\ \bibinfo {pages} {493} (\bibinfo {year}
  {2011})}\BibitemShut {NoStop}%
\bibitem [{\citenamefont {Meyer}\ and\ \citenamefont
  {Bohn}(2008)}]{meyer2008prospects}%
  \BibitemOpen
  \bibfield  {author} {\bibinfo {author} {\bibfnamefont {E.~R.}\ \bibnamefont
  {Meyer}}\ and\ \bibinfo {author} {\bibfnamefont {J.~L.}\ \bibnamefont
  {Bohn}},\ }\bibfield  {title} {\bibinfo {title} {Prospects for an electron
  electric-dipole moment search in metastable {ThO} and {ThF$^+$}},\
  }\href@noop {} {\bibfield  {journal} {\bibinfo  {journal} {Physical Review
  A}\ }\textbf {\bibinfo {volume} {78}},\ \bibinfo {pages} {010502} (\bibinfo
  {year} {2008})}\BibitemShut {NoStop}%
\bibitem [{\citenamefont {Denis}\ \emph {et~al.}(2015)\citenamefont {Denis},
  \citenamefont {N{\o}rby}, \citenamefont {Jensen}, \citenamefont {Gomes},
  \citenamefont {Nayak}, \citenamefont {Knecht},\ and\ \citenamefont
  {Fleig}}]{Denis2015}%
  \BibitemOpen
  \bibfield  {author} {\bibinfo {author} {\bibfnamefont {M.}~\bibnamefont
  {Denis}}, \bibinfo {author} {\bibfnamefont {M.~S.}\ \bibnamefont {N{\o}rby}},
  \bibinfo {author} {\bibfnamefont {H.~J.~A.}\ \bibnamefont {Jensen}}, \bibinfo
  {author} {\bibfnamefont {A.~S.~P.}\ \bibnamefont {Gomes}}, \bibinfo {author}
  {\bibfnamefont {M.~K.}\ \bibnamefont {Nayak}}, \bibinfo {author}
  {\bibfnamefont {S.}~\bibnamefont {Knecht}},\ and\ \bibinfo {author}
  {\bibfnamefont {T.}~\bibnamefont {Fleig}},\ }\bibfield  {title} {\bibinfo
  {title} {Theoretical study on {ThF$^{+}$}, a prospective system in search of
  time-reversal violation},\ }\href@noop {} {\bibfield  {journal} {\bibinfo
  {journal} {New Journal of Physics}\ }\textbf {\bibinfo {volume} {17}},\
  \bibinfo {pages} {043005} (\bibinfo {year} {2015})}\BibitemShut {NoStop}%
\bibitem [{\citenamefont {Skripnikov}\ and\ \citenamefont
  {Titov}(2015{\natexlab{a}})}]{skripnikov2015theoretical}%
  \BibitemOpen
  \bibfield  {author} {\bibinfo {author} {\bibfnamefont {L.~V.}\ \bibnamefont
  {Skripnikov}}\ and\ \bibinfo {author} {\bibfnamefont {A.~V.}\ \bibnamefont
  {Titov}},\ }\bibfield  {title} {\bibinfo {title} {Theoretical study of
  {ThF$^+$} in the search for {T}, {P}-violation effects: Effective state of a
  {Th} atom in {ThF$^+$} and {ThO} compounds},\ }\href@noop {} {\bibfield
  {journal} {\bibinfo  {journal} {Physical Review A}\ }\textbf {\bibinfo
  {volume} {91}},\ \bibinfo {pages} {042504} (\bibinfo {year}
  {2015}{\natexlab{a}})}\BibitemShut {NoStop}%
\bibitem [{\citenamefont {Petrov}\ \emph {et~al.}(2007)\citenamefont {Petrov},
  \citenamefont {Mosyagin}, \citenamefont {Isaev},\ and\ \citenamefont
  {Titov}}]{petrov2007theoretical}%
  \BibitemOpen
  \bibfield  {author} {\bibinfo {author} {\bibfnamefont {A.~N.}\ \bibnamefont
  {Petrov}}, \bibinfo {author} {\bibfnamefont {N.~S.}\ \bibnamefont
  {Mosyagin}}, \bibinfo {author} {\bibfnamefont {T.~A.}\ \bibnamefont
  {Isaev}},\ and\ \bibinfo {author} {\bibfnamefont {A.~V.}\ \bibnamefont
  {Titov}},\ }\bibfield  {title} {\bibinfo {title} {Theoretical study of
  {HfF}$^{+}$ in search of the electron electric dipole moment},\ }\href@noop
  {} {\bibfield  {journal} {\bibinfo  {journal} {Physical Review A}\ }\textbf
  {\bibinfo {volume} {76}},\ \bibinfo {pages} {030501} (\bibinfo {year}
  {2007})}\BibitemShut {NoStop}%
\bibitem [{\citenamefont {Leanhardt}\ \emph {et~al.}(2011)\citenamefont
  {Leanhardt}, \citenamefont {Bohn}, \citenamefont {Loh}, \citenamefont
  {Maletinsky}, \citenamefont {Meyer}, \citenamefont {Sinclair}, \citenamefont
  {Stutz},\ and\ \citenamefont {Cornell}}]{leanhardt2011high}%
  \BibitemOpen
  \bibfield  {author} {\bibinfo {author} {\bibfnamefont {A.~E.}\ \bibnamefont
  {Leanhardt}}, \bibinfo {author} {\bibfnamefont {J.~L.}\ \bibnamefont {Bohn}},
  \bibinfo {author} {\bibfnamefont {H.}~\bibnamefont {Loh}}, \bibinfo {author}
  {\bibfnamefont {P.}~\bibnamefont {Maletinsky}}, \bibinfo {author}
  {\bibfnamefont {E.~R.}\ \bibnamefont {Meyer}}, \bibinfo {author}
  {\bibfnamefont {L.~C.}\ \bibnamefont {Sinclair}}, \bibinfo {author}
  {\bibfnamefont {R.~P.}\ \bibnamefont {Stutz}},\ and\ \bibinfo {author}
  {\bibfnamefont {E.~A.}\ \bibnamefont {Cornell}},\ }\bibfield  {title}
  {\bibinfo {title} {High-resolution spectroscopy on trapped molecular ions in
  rotating electric fields: A new approach for measuring the electron electric
  dipole moment},\ }\href@noop {} {\bibfield  {journal} {\bibinfo  {journal}
  {Journal of Molecular Spectroscopy}\ }\textbf {\bibinfo {volume} {270}},\
  \bibinfo {pages} {1} (\bibinfo {year} {2011})}\BibitemShut {NoStop}%
\bibitem [{\citenamefont {Skripnikov}\ \emph {et~al.}(2013)\citenamefont
  {Skripnikov}, \citenamefont {Petrov},\ and\ \citenamefont
  {Titov}}]{skripnikov2013communication}%
  \BibitemOpen
  \bibfield  {author} {\bibinfo {author} {\bibfnamefont {L.~V.}\ \bibnamefont
  {Skripnikov}}, \bibinfo {author} {\bibfnamefont {A.~N.}\ \bibnamefont
  {Petrov}},\ and\ \bibinfo {author} {\bibfnamefont {A.~V.}\ \bibnamefont
  {Titov}},\ }\href@noop {} {\bibinfo {title} {Communication: Theoretical study
  of {ThO} for the electron electric dipole moment search}} (\bibinfo {year}
  {2013})\BibitemShut {NoStop}%
\bibitem [{\citenamefont {Skripnikov}\ and\ \citenamefont
  {Titov}(2015{\natexlab{b}})}]{skripnikov2015theoreticalThO}%
  \BibitemOpen
  \bibfield  {author} {\bibinfo {author} {\bibfnamefont {L.~V.}\ \bibnamefont
  {Skripnikov}}\ and\ \bibinfo {author} {\bibfnamefont {A.~V.}\ \bibnamefont
  {Titov}},\ }\bibfield  {title} {\bibinfo {title} {Theoretical study of
  thorium monoxide for the electron electric dipole moment search: Electronic
  properties of {$H\,^3\Delta_1$} in {ThO}},\ }\href@noop {} {\bibfield
  {journal} {\bibinfo  {journal} {Journal of Chemical Physics}\ }\textbf
  {\bibinfo {volume} {142}},\ \bibinfo {pages} {024301} (\bibinfo {year}
  {2015}{\natexlab{b}})}\BibitemShut {NoStop}%
\bibitem [{\citenamefont {Baron}\ \emph {et~al.}(2014)\citenamefont {Baron},
  \citenamefont {Campbell}, \citenamefont {DeMille}, \citenamefont {Doyle},
  \citenamefont {Gabrielse}, \citenamefont {Gurevich}, \citenamefont {Hess},
  \citenamefont {Hutzler}, \citenamefont {Kirilov}, \citenamefont {Kozyryev}
  \emph {et~al.}}]{baron2014order}%
  \BibitemOpen
  \bibfield  {author} {\bibinfo {author} {\bibfnamefont {J.}~\bibnamefont
  {Baron}}, \bibinfo {author} {\bibfnamefont {W.~C.}\ \bibnamefont {Campbell}},
  \bibinfo {author} {\bibfnamefont {D.}~\bibnamefont {DeMille}}, \bibinfo
  {author} {\bibfnamefont {J.~M.}\ \bibnamefont {Doyle}}, \bibinfo {author}
  {\bibfnamefont {G.}~\bibnamefont {Gabrielse}}, \bibinfo {author}
  {\bibfnamefont {Y.~V.}\ \bibnamefont {Gurevich}}, \bibinfo {author}
  {\bibfnamefont {P.~W.}\ \bibnamefont {Hess}}, \bibinfo {author}
  {\bibfnamefont {N.~R.}\ \bibnamefont {Hutzler}}, \bibinfo {author}
  {\bibfnamefont {E.}~\bibnamefont {Kirilov}}, \bibinfo {author} {\bibfnamefont
  {I.}~\bibnamefont {Kozyryev}}, \emph {et~al.},\ }\bibfield  {title} {\bibinfo
  {title} {Order of magnitude smaller limit on the electric dipole moment of
  the electron},\ }\href@noop {} {\bibfield  {journal} {\bibinfo  {journal}
  {Science}\ }\textbf {\bibinfo {volume} {343}},\ \bibinfo {pages} {269}
  (\bibinfo {year} {2014})}\BibitemShut {NoStop}%
\bibitem [{\citenamefont {Zhou}\ \emph {et~al.}(2020)\citenamefont {Zhou},
  \citenamefont {Shagam}, \citenamefont {Cairncross}, \citenamefont {Ng},
  \citenamefont {Roussy}, \citenamefont {Grogan}, \citenamefont {Boyce},
  \citenamefont {Vigil}, \citenamefont {Pettine}, \citenamefont {Zelevinsky}
  \emph {et~al.}}]{zhou2020second}%
  \BibitemOpen
  \bibfield  {author} {\bibinfo {author} {\bibfnamefont {Y.}~\bibnamefont
  {Zhou}}, \bibinfo {author} {\bibfnamefont {Y.}~\bibnamefont {Shagam}},
  \bibinfo {author} {\bibfnamefont {W.~B.}\ \bibnamefont {Cairncross}},
  \bibinfo {author} {\bibfnamefont {K.~B.}\ \bibnamefont {Ng}}, \bibinfo
  {author} {\bibfnamefont {T.~S.}\ \bibnamefont {Roussy}}, \bibinfo {author}
  {\bibfnamefont {T.}~\bibnamefont {Grogan}}, \bibinfo {author} {\bibfnamefont
  {K.}~\bibnamefont {Boyce}}, \bibinfo {author} {\bibfnamefont
  {A.}~\bibnamefont {Vigil}}, \bibinfo {author} {\bibfnamefont
  {M.}~\bibnamefont {Pettine}}, \bibinfo {author} {\bibfnamefont
  {T.}~\bibnamefont {Zelevinsky}}, \emph {et~al.},\ }\bibfield  {title}
  {\bibinfo {title} {Second-scale coherence measured at the quantum projection
  noise limit with hundreds of molecular ions},\ }\href@noop {} {\bibfield
  {journal} {\bibinfo  {journal} {Physical Review Letters}\ }\textbf {\bibinfo
  {volume} {124}},\ \bibinfo {pages} {053201} (\bibinfo {year}
  {2020})}\BibitemShut {NoStop}%
\bibitem [{\citenamefont {Gresh}\ \emph {et~al.}(2016)\citenamefont {Gresh},
  \citenamefont {Cossel}, \citenamefont {Zhou}, \citenamefont {Ye},\ and\
  \citenamefont {Cornell}}]{gresh2016broadband}%
  \BibitemOpen
  \bibfield  {author} {\bibinfo {author} {\bibfnamefont {D.~N.}\ \bibnamefont
  {Gresh}}, \bibinfo {author} {\bibfnamefont {K.~C.}\ \bibnamefont {Cossel}},
  \bibinfo {author} {\bibfnamefont {Y.}~\bibnamefont {Zhou}}, \bibinfo {author}
  {\bibfnamefont {J.}~\bibnamefont {Ye}},\ and\ \bibinfo {author}
  {\bibfnamefont {E.~A.}\ \bibnamefont {Cornell}},\ }\bibfield  {title}
  {\bibinfo {title} {Broadband velocity modulation spectroscopy of {ThF}$^+$
  for use in a measurement of the electron electric dipole moment},\
  }\href@noop {} {\bibfield  {journal} {\bibinfo  {journal} {Journal of
  Molecular Spectroscopy}\ }\textbf {\bibinfo {volume} {319}},\ \bibinfo
  {pages} {1} (\bibinfo {year} {2016})}\BibitemShut {NoStop}%
\bibitem [{\citenamefont {Zhou}\ \emph {et~al.}(2019)\citenamefont {Zhou},
  \citenamefont {Ng}, \citenamefont {Cheng}, \citenamefont {Gresh},
  \citenamefont {Field}, \citenamefont {Ye},\ and\ \citenamefont
  {Cornell}}]{zhou2019visible}%
  \BibitemOpen
  \bibfield  {author} {\bibinfo {author} {\bibfnamefont {Y.}~\bibnamefont
  {Zhou}}, \bibinfo {author} {\bibfnamefont {K.~B.}\ \bibnamefont {Ng}},
  \bibinfo {author} {\bibfnamefont {L.}~\bibnamefont {Cheng}}, \bibinfo
  {author} {\bibfnamefont {D.~N.}\ \bibnamefont {Gresh}}, \bibinfo {author}
  {\bibfnamefont {R.~W.}\ \bibnamefont {Field}}, \bibinfo {author}
  {\bibfnamefont {J.}~\bibnamefont {Ye}},\ and\ \bibinfo {author}
  {\bibfnamefont {E.~A.}\ \bibnamefont {Cornell}},\ }\bibfield  {title}
  {\bibinfo {title} {Visible and ultraviolet laser spectroscopy of {ThF}},\
  }\href@noop {} {\bibfield  {journal} {\bibinfo  {journal} {Journal of
  Molecular Spectroscopy}\ }\textbf {\bibinfo {volume} {358}},\ \bibinfo
  {pages} {1} (\bibinfo {year} {2019})}\BibitemShut {NoStop}%
\bibitem [{\citenamefont {Heaven}\ \emph {et~al.}(2014)\citenamefont {Heaven},
  \citenamefont {Barker},\ and\ \citenamefont
  {Antonov}}]{heaven2014spectroscopy}%
  \BibitemOpen
  \bibfield  {author} {\bibinfo {author} {\bibfnamefont {M.~C.}\ \bibnamefont
  {Heaven}}, \bibinfo {author} {\bibfnamefont {B.~J.}\ \bibnamefont {Barker}},\
  and\ \bibinfo {author} {\bibfnamefont {I.~O.}\ \bibnamefont {Antonov}},\
  }\bibfield  {title} {\bibinfo {title} {Spectroscopy and structure of the
  simplest actinide bonds},\ }\href@noop {} {\bibfield  {journal} {\bibinfo
  {journal} {The Journal of Physical Chemistry A}\ }\textbf {\bibinfo {volume}
  {118}},\ \bibinfo {pages} {10867} (\bibinfo {year} {2014})}\BibitemShut
  {NoStop}%
\bibitem [{\citenamefont {Barker}\ \emph {et~al.}(2012)\citenamefont {Barker},
  \citenamefont {Antonov}, \citenamefont {Heaven},\ and\ \citenamefont
  {Peterson}}]{barker2012spectroscopic}%
  \BibitemOpen
  \bibfield  {author} {\bibinfo {author} {\bibfnamefont {B.~J.}\ \bibnamefont
  {Barker}}, \bibinfo {author} {\bibfnamefont {I.~O.}\ \bibnamefont {Antonov}},
  \bibinfo {author} {\bibfnamefont {M.~C.}\ \bibnamefont {Heaven}},\ and\
  \bibinfo {author} {\bibfnamefont {K.~A.}\ \bibnamefont {Peterson}},\
  }\bibfield  {title} {\bibinfo {title} {Spectroscopic investigations of {ThF}
  and {ThF}$^+$},\ }\href@noop {} {\bibfield  {journal} {\bibinfo  {journal}
  {Journal of Chemical Physics}\ }\textbf {\bibinfo {volume} {136}},\ \bibinfo
  {pages} {104305} (\bibinfo {year} {2012})}\BibitemShut {NoStop}%
\bibitem [{\citenamefont {Shagam}\ \emph {et~al.}(2020)\citenamefont {Shagam},
  \citenamefont {Cairncross}, \citenamefont {Roussy}, \citenamefont {Zhou},
  \citenamefont {Ng}, \citenamefont {Gresh}, \citenamefont {Grogan},
  \citenamefont {Ye},\ and\ \citenamefont {Cornell}}]{shagam2020continuous}%
  \BibitemOpen
  \bibfield  {author} {\bibinfo {author} {\bibfnamefont {Y.}~\bibnamefont
  {Shagam}}, \bibinfo {author} {\bibfnamefont {W.~B.}\ \bibnamefont
  {Cairncross}}, \bibinfo {author} {\bibfnamefont {T.~S.}\ \bibnamefont
  {Roussy}}, \bibinfo {author} {\bibfnamefont {Y.}~\bibnamefont {Zhou}},
  \bibinfo {author} {\bibfnamefont {K.~B.}\ \bibnamefont {Ng}}, \bibinfo
  {author} {\bibfnamefont {D.~N.}\ \bibnamefont {Gresh}}, \bibinfo {author}
  {\bibfnamefont {T.}~\bibnamefont {Grogan}}, \bibinfo {author} {\bibfnamefont
  {J.}~\bibnamefont {Ye}},\ and\ \bibinfo {author} {\bibfnamefont {E.~A.}\
  \bibnamefont {Cornell}},\ }\bibfield  {title} {\bibinfo {title} {Continuous
  temporal ion detection combined with time-gated imaging: normalization over a
  large dynamic range},\ }\href@noop {} {\bibfield  {journal} {\bibinfo
  {journal} {Journal of Molecular Spectroscopy}\ ,\ \bibinfo {pages} {111257}}
  (\bibinfo {year} {2020})}\BibitemShut {NoStop}%
\bibitem [{\citenamefont {Loh}\ \emph {et~al.}(2011)\citenamefont {Loh},
  \citenamefont {Wang}, \citenamefont {Grau}, \citenamefont {Yahn},
  \citenamefont {Field}, \citenamefont {Greene},\ and\ \citenamefont
  {Cornell}}]{loh2011laser}%
  \BibitemOpen
  \bibfield  {author} {\bibinfo {author} {\bibfnamefont {H.}~\bibnamefont
  {Loh}}, \bibinfo {author} {\bibfnamefont {J.}~\bibnamefont {Wang}}, \bibinfo
  {author} {\bibfnamefont {M.}~\bibnamefont {Grau}}, \bibinfo {author}
  {\bibfnamefont {T.~S.}\ \bibnamefont {Yahn}}, \bibinfo {author}
  {\bibfnamefont {R.~W.}\ \bibnamefont {Field}}, \bibinfo {author}
  {\bibfnamefont {C.~H.}\ \bibnamefont {Greene}},\ and\ \bibinfo {author}
  {\bibfnamefont {E.~A.}\ \bibnamefont {Cornell}},\ }\bibfield  {title}
  {\bibinfo {title} {Laser-induced fluorescence studies of {HfF$^{+}$} produced
  by autoionization},\ }\href@noop {} {\bibfield  {journal} {\bibinfo
  {journal} {Journal of Chemical Physics}\ }\textbf {\bibinfo {volume} {135}},\
  \bibinfo {pages} {154308} (\bibinfo {year} {2011})}\BibitemShut {NoStop}%
\bibitem [{\citenamefont {Loh}\ \emph {et~al.}(2012)\citenamefont {Loh},
  \citenamefont {Stutz}, \citenamefont {Yahn}, \citenamefont {Looser},
  \citenamefont {Field},\ and\ \citenamefont {Cornell}}]{loh2012rempi}%
  \BibitemOpen
  \bibfield  {author} {\bibinfo {author} {\bibfnamefont {H.}~\bibnamefont
  {Loh}}, \bibinfo {author} {\bibfnamefont {R.~P.}\ \bibnamefont {Stutz}},
  \bibinfo {author} {\bibfnamefont {T.~S.}\ \bibnamefont {Yahn}}, \bibinfo
  {author} {\bibfnamefont {H.}~\bibnamefont {Looser}}, \bibinfo {author}
  {\bibfnamefont {R.~W.}\ \bibnamefont {Field}},\ and\ \bibinfo {author}
  {\bibfnamefont {E.~A.}\ \bibnamefont {Cornell}},\ }\bibfield  {title}
  {\bibinfo {title} {Rempi spectroscopy of {HfF}$^+$},\ }\href@noop {}
  {\bibfield  {journal} {\bibinfo  {journal} {Journal of Molecular
  Spectroscopy}\ }\textbf {\bibinfo {volume} {276}},\ \bibinfo {pages} {49}
  (\bibinfo {year} {2012})}\BibitemShut {NoStop}%
\bibitem [{\citenamefont {Ni}\ \emph {et~al.}(2014)\citenamefont {Ni},
  \citenamefont {Loh}, \citenamefont {Grau}, \citenamefont {Cossel},
  \citenamefont {Ye},\ and\ \citenamefont {Cornell}}]{ni2014state}%
  \BibitemOpen
  \bibfield  {author} {\bibinfo {author} {\bibfnamefont {K.-K.}\ \bibnamefont
  {Ni}}, \bibinfo {author} {\bibfnamefont {H.}~\bibnamefont {Loh}}, \bibinfo
  {author} {\bibfnamefont {M.}~\bibnamefont {Grau}}, \bibinfo {author}
  {\bibfnamefont {K.~C.}\ \bibnamefont {Cossel}}, \bibinfo {author}
  {\bibfnamefont {J.}~\bibnamefont {Ye}},\ and\ \bibinfo {author}
  {\bibfnamefont {E.~A.}\ \bibnamefont {Cornell}},\ }\bibfield  {title}
  {\bibinfo {title} {State-specific detection of trapped {HfF$^+$} by
  photodissociation},\ }\href@noop {} {\bibfield  {journal} {\bibinfo
  {journal} {Journal of Molecular Spectroscopy}\ }\textbf {\bibinfo {volume}
  {300}},\ \bibinfo {pages} {12} (\bibinfo {year} {2014})}\BibitemShut
  {NoStop}%
\bibitem [{\citenamefont {Loh}\ \emph {et~al.}(2013)\citenamefont {Loh},
  \citenamefont {Cossel}, \citenamefont {Grau}, \citenamefont {Ni},
  \citenamefont {Meyer}, \citenamefont {Bohn}, \citenamefont {Ye},\ and\
  \citenamefont {Cornell}}]{loh2013precision}%
  \BibitemOpen
  \bibfield  {author} {\bibinfo {author} {\bibfnamefont {H.}~\bibnamefont
  {Loh}}, \bibinfo {author} {\bibfnamefont {K.~C.}\ \bibnamefont {Cossel}},
  \bibinfo {author} {\bibfnamefont {M.}~\bibnamefont {Grau}}, \bibinfo {author}
  {\bibfnamefont {K.-K.}\ \bibnamefont {Ni}}, \bibinfo {author} {\bibfnamefont
  {E.~R.}\ \bibnamefont {Meyer}}, \bibinfo {author} {\bibfnamefont {J.~L.}\
  \bibnamefont {Bohn}}, \bibinfo {author} {\bibfnamefont {J.}~\bibnamefont
  {Ye}},\ and\ \bibinfo {author} {\bibfnamefont {E.~A.}\ \bibnamefont
  {Cornell}},\ }\bibfield  {title} {\bibinfo {title} {Precision spectroscopy of
  polarized molecules in an ion trap},\ }\href@noop {} {\bibfield  {journal}
  {\bibinfo  {journal} {Science}\ }\textbf {\bibinfo {volume} {342}},\ \bibinfo
  {pages} {1220} (\bibinfo {year} {2013})}\BibitemShut {NoStop}%
\bibitem [{\citenamefont {Raghavachari}\ \emph {et~al.}(1989)\citenamefont
  {Raghavachari}, \citenamefont {Trucks}, \citenamefont {Pople},\ and\
  \citenamefont {Head-Gordon}}]{Pople1989}%
  \BibitemOpen
  \bibfield  {author} {\bibinfo {author} {\bibfnamefont {K.}~\bibnamefont
  {Raghavachari}}, \bibinfo {author} {\bibfnamefont {G.~W.}\ \bibnamefont
  {Trucks}}, \bibinfo {author} {\bibfnamefont {J.~A.}\ \bibnamefont {Pople}},\
  and\ \bibinfo {author} {\bibfnamefont {M.}~\bibnamefont {Head-Gordon}},\
  }\bibfield  {title} {\bibinfo {title} {A fifth-order perturbation comparison
  of electron correlation theories},\ }\href@noop {} {\bibfield  {journal}
  {\bibinfo  {journal} {Chemical Physics Letters}\ }\textbf {\bibinfo {volume}
  {157}},\ \bibinfo {pages} {479} (\bibinfo {year} {1989})}\BibitemShut
  {NoStop}%
\bibitem [{\citenamefont {Petrov}\ \emph {et~al.}(2017)\citenamefont {Petrov},
  \citenamefont {Skripnikov},\ and\ \citenamefont {Titov}}]{petrov2017zeeman}%
  \BibitemOpen
  \bibfield  {author} {\bibinfo {author} {\bibfnamefont {A.~N.}\ \bibnamefont
  {Petrov}}, \bibinfo {author} {\bibfnamefont {L.~V.}\ \bibnamefont
  {Skripnikov}},\ and\ \bibinfo {author} {\bibfnamefont {A.~V.}\ \bibnamefont
  {Titov}},\ }\bibfield  {title} {\bibinfo {title} {Zeeman interaction in the
  {$^3\Delta_1$} state of {HfF$^+$} to search for the electron electric dipole
  moment},\ }\href@noop {} {\bibfield  {journal} {\bibinfo  {journal} {Physical
  Review A}\ }\textbf {\bibinfo {volume} {96}},\ \bibinfo {pages} {022508}
  (\bibinfo {year} {2017})}\BibitemShut {NoStop}%
\bibitem [{\citenamefont {Dyall}(1997)}]{Dyall1997}%
  \BibitemOpen
  \bibfield  {author} {\bibinfo {author} {\bibfnamefont {K.~G.}\ \bibnamefont
  {Dyall}},\ }\bibfield  {title} {\bibinfo {title} {{Interfacing relativistic
  and nonrelativistic methods. I. Normalized elimination of the small component
  in the modified Dirac equation}},\ }\href@noop {} {\bibfield  {journal}
  {\bibinfo  {journal} {Journal of Chemical Physics}\ }\textbf {\bibinfo
  {volume} {106}},\ \bibinfo {pages} {9618} (\bibinfo {year}
  {1997})}\BibitemShut {NoStop}%
\bibitem [{\citenamefont {Kutzelnigg}\ and\ \citenamefont {Liu}(2005)}]{XQR}%
  \BibitemOpen
  \bibfield  {author} {\bibinfo {author} {\bibfnamefont {W.}~\bibnamefont
  {Kutzelnigg}}\ and\ \bibinfo {author} {\bibfnamefont {W.}~\bibnamefont
  {Liu}},\ }\bibfield  {title} {\bibinfo {title} {Quasirelativistic theory
  equivalent to fully relativistic theory},\ }\href@noop {} {\bibfield
  {journal} {\bibinfo  {journal} {Journal of Chemical Physucs}\ }\textbf
  {\bibinfo {volume} {123}},\ \bibinfo {pages} {241102} (\bibinfo {year}
  {2005})}\BibitemShut {NoStop}%
\bibitem [{\citenamefont {Liu}\ and\ \citenamefont {Cheng}(2018)}]{Liu2018}%
  \BibitemOpen
  \bibfield  {author} {\bibinfo {author} {\bibfnamefont {J.}~\bibnamefont
  {Liu}}\ and\ \bibinfo {author} {\bibfnamefont {L.}~\bibnamefont {Cheng}},\
  }\bibfield  {title} {\bibinfo {title} {An atomic mean-field spin-orbit
  approach within exact two-component theory for a non-perturbative treatment
  of spin-orbit coupling},\ }\href@noop {} {\bibfield  {journal} {\bibinfo
  {journal} {Journal of Chemical Physics}\ }\textbf {\bibinfo {volume} {148}},\
  \bibinfo {pages} {144108} (\bibinfo {year} {2018})}\BibitemShut {NoStop}%
\bibitem [{\citenamefont {Matthews}\ \emph {et~al.}(2020)\citenamefont
  {Matthews}, \citenamefont {Cheng}, \citenamefont {Harding}, \citenamefont
  {Lipparini}, \citenamefont {Stopkowicz}, \citenamefont {Jagau}, \citenamefont
  {Szalay}, \citenamefont {Gauss},\ and\ \citenamefont
  {Stanton}}]{Matthews20CFOUR}%
  \BibitemOpen
  \bibfield  {author} {\bibinfo {author} {\bibfnamefont {D.~A.}\ \bibnamefont
  {Matthews}}, \bibinfo {author} {\bibfnamefont {L.}~\bibnamefont {Cheng}},
  \bibinfo {author} {\bibfnamefont {M.~E.}\ \bibnamefont {Harding}}, \bibinfo
  {author} {\bibfnamefont {F.}~\bibnamefont {Lipparini}}, \bibinfo {author}
  {\bibfnamefont {S.}~\bibnamefont {Stopkowicz}}, \bibinfo {author}
  {\bibfnamefont {T.-C.}\ \bibnamefont {Jagau}}, \bibinfo {author}
  {\bibfnamefont {P.~G.}\ \bibnamefont {Szalay}}, \bibinfo {author}
  {\bibfnamefont {J.}~\bibnamefont {Gauss}},\ and\ \bibinfo {author}
  {\bibfnamefont {J.~F.}\ \bibnamefont {Stanton}},\ }\bibfield  {title}
  {\bibinfo {title} {{Coupled-cluster techniques for computational chemistry:
  The CFOUR program package}},\ }\href@noop {} {\bibfield  {journal} {\bibinfo
  {journal} {J. Chem. Phys.}\ }\textbf {\bibinfo {volume} {152}},\ \bibinfo
  {pages} {214108} (\bibinfo {year} {2020})}\BibitemShut {NoStop}%
\bibitem [{\citenamefont {Stanton}\ \emph {et~al.}()\citenamefont {Stanton},
  \citenamefont {Gauss}, \citenamefont {Cheng}, \citenamefont {Harding},
  \citenamefont {Matthews},\ and\ \citenamefont {Szalay}}]{cfour}%
  \BibitemOpen
  \bibfield  {author} {\bibinfo {author} {\bibfnamefont {J.~F.}\ \bibnamefont
  {Stanton}}, \bibinfo {author} {\bibfnamefont {J.}~\bibnamefont {Gauss}},
  \bibinfo {author} {\bibfnamefont {L.}~\bibnamefont {Cheng}}, \bibinfo
  {author} {\bibfnamefont {M.~E.}\ \bibnamefont {Harding}}, \bibinfo {author}
  {\bibfnamefont {D.~A.}\ \bibnamefont {Matthews}},\ and\ \bibinfo {author}
  {\bibfnamefont {P.~G.}\ \bibnamefont {Szalay}},\ }\href@noop {} {\bibinfo
  {title} {{CFOUR, Coupled-Cluster techniques for Computational Chemistry, a
  quantum-chemical program package}}},\ \bibinfo {note} {{W}ith contributions
  from {A}.{A}. {A}uer, {A}. {A}sthana, {R}.{J}. {B}artlett, {U}. {B}enedikt,
  {C}. {B}erger, {D}.{E}. {B}ernholdt, {S.} {B}laschke, {Y}. {J}. {B}omble,
  {S.} {B}urger, {O}. {C}hristiansen, {D.} Datta, {F}. Engel, {R}. Faber, {J.}
  {G}reiner, {M}. {H}eckert, {O}. {H}eun, {M}. Hilgenberg, {C}. {H}uber,
  {T}.-{C}. {J}agau, {D}. {J}onsson, {J}. {J}us{\'e}lius, {T}. Kirsch, {K}.
  {K}lein, {G}.{M.} Kopper{W}.{J}. {L}auderdale, {F}. {L}ipparini, {J}. {L}iu,
  {T}. {M}etzroth, {L}.{A}. {M}{\"u}ck, {D}.{P}. {O}'{N}eill, {T.} {N}ottoli,
  {D}.{R}. {P}rice, {E}. {P}rochnow, {C}. {P}uzzarini, {K}. {R}uud, {F}.
  {S}chiffmann, {W}. {S}chwalbach, {C}. {S}immons, {S}. {S}topkowicz, {A}.
  {T}ajti, {J}. {V}{\'a}zquez, {F}. {W}ang, {J}.{D}. {W}atts and the integral
  packages {MOLECULE} ({J}. {A}lml{\"o}f and {P}.{R}. {T}aylor), {PROPS}
  ({P}.{R}. {T}aylor), {ABACUS} ({T}. {H}elgaker, {H}.{J}. {A}a. {J}ensen, {P}.
  {J}{\o}rgensen, and {J}. {O}lsen), and {ECP} routines by {A}. {V}. {M}itin
  and {C}. van {W}{\"u}llen. {F}or the current version, see
  http://www.cfour.de.}\BibitemShut {Stop}%
\bibitem [{\citenamefont {Kutzelnigg}(2003)}]{UT}%
  \BibitemOpen
  \bibfield  {author} {\bibinfo {author} {\bibfnamefont {W.}~\bibnamefont
  {Kutzelnigg}},\ }\bibfield  {title} {\bibinfo {title} {Diamagnetism in
  relativistic theory},\ }\href@noop {} {\bibfield  {journal} {\bibinfo
  {journal} {Physical Review A}\ }\textbf {\bibinfo {volume} {67}},\ \bibinfo
  {pages} {032109} (\bibinfo {year} {2003})}\BibitemShut {NoStop}%
\bibitem [{\citenamefont {Roos}\ \emph {et~al.}(2005)\citenamefont {Roos},
  \citenamefont {Lindh}, \citenamefont {Malmqvist}, \citenamefont {Veryazov},\
  and\ \citenamefont {Widmark}}]{ROOS2005}%
  \BibitemOpen
  \bibfield  {author} {\bibinfo {author} {\bibfnamefont {B.~O.}\ \bibnamefont
  {Roos}}, \bibinfo {author} {\bibfnamefont {R.}~\bibnamefont {Lindh}},
  \bibinfo {author} {\bibfnamefont {P.-{\AA}.}\ \bibnamefont {Malmqvist}},
  \bibinfo {author} {\bibfnamefont {V.}~\bibnamefont {Veryazov}},\ and\
  \bibinfo {author} {\bibfnamefont {P.-O.}\ \bibnamefont {Widmark}},\
  }\bibfield  {title} {\bibinfo {title} {New relativistic {ANO} basis sets for
  actinide atoms},\ }\href@noop {} {\bibfield  {journal} {\bibinfo  {journal}
  {Chemical Physics Letters}\ }\textbf {\bibinfo {volume} {409}},\ \bibinfo
  {pages} {295} (\bibinfo {year} {2005})}\BibitemShut {NoStop}%
\bibitem [{\citenamefont {{Faegri Jr}}(2001)}]{Faegri2001}%
  \BibitemOpen
  \bibfield  {author} {\bibinfo {author} {\bibfnamefont {K.}~\bibnamefont
  {{Faegri Jr}}},\ }\bibfield  {title} {\bibinfo {title} {Relativistic gaussian
  basis sets for the elements {K-Uuo}},\ }\href@noop {} {\bibfield  {journal}
  {\bibinfo  {journal} {Theoretical Chemistry Accounts}\ }\textbf {\bibinfo
  {volume} {105}},\ \bibinfo {pages} {252} (\bibinfo {year}
  {2001})}\BibitemShut {NoStop}%
\bibitem [{\citenamefont {Feng}\ and\ \citenamefont
  {Peterson}(2017)}]{Th-pwcvtz}%
  \BibitemOpen
  \bibfield  {author} {\bibinfo {author} {\bibfnamefont {R.}~\bibnamefont
  {Feng}}\ and\ \bibinfo {author} {\bibfnamefont {K.~A.}\ \bibnamefont
  {Peterson}},\ }\bibfield  {title} {\bibinfo {title} {{Correlation consistent
  basis sets for actinides. II. The atoms Ac and Np–Lr}},\ }\href@noop {}
  {\bibfield  {journal} {\bibinfo  {journal} {Journal of Chemical Physics}\
  }\textbf {\bibinfo {volume} {147}},\ \bibinfo {pages} {84108} (\bibinfo
  {year} {2017})}\BibitemShut {NoStop}%
\bibitem [{\citenamefont {Petrov}\ \emph {et~al.}(2014)\citenamefont {Petrov},
  \citenamefont {Skripnikov}, \citenamefont {Titov}, \citenamefont {Hutzler},
  \citenamefont {Hess}, \citenamefont {O'Leary}, \citenamefont {Spaun},
  \citenamefont {DeMille}, \citenamefont {Gabrielse},\ and\ \citenamefont
  {Doyle}}]{ThOgf}%
  \BibitemOpen
  \bibfield  {author} {\bibinfo {author} {\bibfnamefont {A.}~\bibnamefont
  {Petrov}}, \bibinfo {author} {\bibfnamefont {L.}~\bibnamefont {Skripnikov}},
  \bibinfo {author} {\bibfnamefont {A.}~\bibnamefont {Titov}}, \bibinfo
  {author} {\bibfnamefont {N.~R.}\ \bibnamefont {Hutzler}}, \bibinfo {author}
  {\bibfnamefont {P.}~\bibnamefont {Hess}}, \bibinfo {author} {\bibfnamefont
  {B.}~\bibnamefont {O'Leary}}, \bibinfo {author} {\bibfnamefont
  {B.}~\bibnamefont {Spaun}}, \bibinfo {author} {\bibfnamefont
  {D.}~\bibnamefont {DeMille}}, \bibinfo {author} {\bibfnamefont
  {G.}~\bibnamefont {Gabrielse}},\ and\ \bibinfo {author} {\bibfnamefont
  {J.~M.}\ \bibnamefont {Doyle}},\ }\bibfield  {title} {\bibinfo {title}
  {Zeeman interaction in {ThO} {$H\,^3\Delta_1$} for the electron
  electric-dipole-moment search},\ }\href@noop {} {\bibfield  {journal}
  {\bibinfo  {journal} {Physical Review A}\ }\textbf {\bibinfo {volume} {89}},\
  \bibinfo {pages} {062505} (\bibinfo {year} {2014})}\BibitemShut {NoStop}%
\end{thebibliography}%


%

\end{document}